\documentclass[twocolumn,english,aps,prb,groupedaddress,showpacs,superscriptaddress,nobibnotes]{revtex4-1}
\usepackage{amsmath,amssymb,amsthm}
\usepackage{newtxmath}
\usepackage[T1]{fontenc}
\usepackage[latin9]{inputenc}
\usepackage{geometry}
\usepackage{color}
\usepackage{verbatim}
\usepackage{esint}
\usepackage{soul}
\usepackage[normalem]{ulem}

\makeatletter
\usepackage{caption}
\usepackage{subcaption}
\usepackage[dvipsnames,svgnames,x11names]{xcolor} 
\usepackage[pdftex]{hyperref}
\hypersetup{
    breaklinks=true,
    colorlinks=true,
    linkcolor=MidnightBlue,
    urlcolor=NavyBlue,
    citecolor=Blue,
    pdftitle={Linking infinite bond-dimension matrix product states with frustration-free Hamiltonians},
    pdfauthor={M. Schossler, L. Chen, A. Seidel},
}

\usepackage{wrapfig}
\usepackage{braket}
\usepackage{array}
\usepackage{relsize}
\usepackage{indentfirst}
\usepackage{simplewick}
\usepackage{mathrsfs}

\usepackage{newtxtext}
\usepackage{mleftright}
\mleftright
\renewcommand{\[}{\begin{equation}}
    \renewcommand{\]}{\end{equation}}

\usepackage{lipsum}
\def\Eq#1{Eq. {\eqref{#1}}}

\usepackage{orcidlink}

\makeatother

\usepackage{babel}
\begin{document}
\title{Linking infinite bond-dimension matrix product states with frustration-free Hamiltonians}
\author{Matheus Schossler\,\orcidlink{0000-0002-4681-059X}}
\email{mschossler@wustl.edu}

\affiliation{Department of Physics, Washington University in St. Louis, 1 Brookings
Dr., St. Louis MO 63130, USA}
\author{Li Chen\,\orcidlink{0000-0002-4870-9065}}
\affiliation{College of Physics and Electronic Science, Hubei Normal University,
Huangshi 435002, China}
\author{Alexander Seidel\,\orcidlink{0000-0002-7008-7063}}
\email{seidel@physics.wustl.edu}
\affiliation{Department of Physics, Washington University in St. Louis, 1 Brookings
Dr., St. Louis MO 63130, USA}
\affiliation{Technical University of Munich, TUM School of Natural Sciences, Physics Department, 85748 Garching, Germany}
\affiliation{Munich Center for Quantum Science and Technology (MCQST), Schellingstr. 4, 80799 M{\"u}nchen, Germany}
\date{\today}

\begin{abstract}

The study of frustration-free Hamiltonians and their relation to finite bond dimension matrix-product-states (MPS) has a long tradition. However, fractional quantum Hall states do not quite fit into this theme, since the known MPS representations of their ground states have infinite bond dimensions, which considerably obscures the relations between such MPS representations and the existence of frustration-free parent Hamiltonians. This is related to the fact that the latter necessarily are of infinite range in the orbital basis. Here we present a Theorem taylored to establishing the existence of frustration free parent Hamiltonians in such a context. We explicitly demonstrate the utility of this Theorem in the context of non-Abelian Moore-Read fractional quantum Hall states, but argue the applicability of this Theorem to transcend considerably beyond the realm of conformal-field-theory-derived matrix product states, or quasi-one-dimensional Hilbert spaces.

\end{abstract}

\maketitle

\section{Introduction }

The theory of fractional quantum Hall (FQH) effect is widely regarded as a theory of beautiful wave functions that are linked rather directly to effective quantum field theory descriptions both for the bulk of the system and for the closely related edge. The close link between microscopic wave function and effective theory can be brought about through powerful mappings and conjectures, such as the Moore-Read (MR) conjecture \cite{Moore1991}.
Given the great success of such mappings between model wave function and effective theory in the exploration of possible phases in the FQH regime, it remains perhaps somewhat under-appreciated that in specific cases, model Hamiltonians can serve to significantly further corroborate 
the universal physics of a given wave function description. 
The fact that such model Hamiltonians are less credited for the enormous success of the theoretical description of FQH states may be due to the fact that known instances were largely limited to a subset of wave functions to which Moore-Read type arguments can be directly applied: They are lowest Landau level (LLL), holomorphic wave functions that can be obtained as conformal blocks in some associated rational conformal field theory. Originally, thus, well-studied parent Hamiltonians in the field stabilized model wave functions whose physics are well under control by the MR conjecture.
More recently, however, it has been demonstrated that the class of FQH wave functions whose long-distance physics can be fully exposed by a systematic study of zero mode spaces of accompanying parent Hamiltonians is considerably larger than previously thought. It contains, for example, model Hamiltonians\cite{Bandyopadhyay2020} for the entire positive Jain sequence, as well as non-Abelian parton-like wave functions.\cite{Jain1989a,Jain1989,Jain1990, Jain1990a,jainbook,compositeReview,Wen1992, Bandyopadhyay2018, Cruise2023, Ahari2022}
This recent progress was made possible not merely by the identification of appropriate parent Hamiltonians, but also by the development of new techniques to rigorously study their zero mode spaces. These developments were made necessary by the presence of higher Landau level degrees of freedom that destroy the holomorphic dependence of the wave function on position variables. The latter is the reason why traditional parent Hamiltonians in the LLL allow rigorous exploration of their zero mode spaces by translating the problem into the search for symmetric polynomials with certain additional clustering conditions. In the more general cases recently studied, this connection with symmetric polynomials is lacking. The recent forays into the rigorous study of FQH parent Hamiltonians of mixed Landau level states have shown that the desirable properties of these Hamiltonians, namely, an analytically accessible ``topological'' zero mode space, 
are in no way tied to underlying symmetric polynomials or limited to the applicability of the ``polynomial-techniques'' traditionally used to establish these spaces. Indeed, it was only the abandoning of these techniques that brought into focus a potentially much larger class of solvable Hamiltonians sharing similar properties. The characteristic of these new techniques, which are equally applicable in the LLL and in the broader context, is the fact that they put greater emphasis on the second-quantized representation of FQH wave functions.

Special parent Hamiltonians of the kind considered here have the property that they divide the Hilbert space into a finite energy subspace (with energies ideally bounded from below by a gap) and a ``topological'' zero energy (``zero modes'') subspace. The latter contains all the universal physics of the underlying quantum Hall state. This is true both for the bulk physics as well as the closely related edge theory. In the case of the bulk, the zero mode space contains localized quasi-hole excitations whose holonomies encode the exchange statistics that characterize the low-energy physics in the bulk, and are well-suited for applications as protected qubits in topological quantum computation \cite{Kitaev2006,Nayak2008}.  Localized quasi-hole states are not angular momentum eigenstates, but can be expanded in an angular-momentum zero mode eigenbasis. Such an eigenbasis will be in one-to-one correspondence with states in the edge conformal field theory (CFT) \cite{Read2009a}. More concretely, the number of zero modes at given angular momentum relative to the incompressible state matches precisely the number of modes at the corresponding level in the edge CFT. A confining potential proportional to total angular momentum then renders {\em small angular momentum} $\Delta L$ (relative to the incompressible state) zero modes the physical gapless edge excitations of the system, whereas quasi-holes in the bulk can be thought of as being formed by ``high energy modes'' in the edge theory. The point is that up to angular momenta $\delta L\sim$ particle number $N$ (where zero modes already represent bulk excitations), the 
number of zero modes of the microscopic model at given $\Delta L$ exactly matches the number of modes at level $\Delta L$ in the effective edge theory.
In such models, thus, whenever the zero mode structure is under control, a large amount of field-theoretic data can be obtained. (The holonomies mentioned above, while also encoded in the zero modes, are very challenging to calculate directly,\cite{Arovas1984, Read2009, Bonderson2011}
but can be simplified considerably with some assumptions\cite{Flavin2011}.)

It is well-known that when presented in the orbital (angular momentum) basis, i.e., in second quantization, 
fractional quantum Hall parent Hamiltonians become one-dimensional lattice models \cite{Lee2004, Seidel2005}. The long-range character of the interaction in this {presentation}  renders the rigorous study of zero mode spaces non-trivial. It is for this reason that historically, such models were studied from as first-quantized models in two spatial dimensions, even though such a perspective somewhat obscures the relevant dynamical degrees of freedom \cite{Haldane2011}. {
The guiding-center degrees of freedom fully encode the topological quantum order. The first clue to this realization was the observation that rotational invariance is not necessary for the FQH effect \cite{Prange1990}. Subsequent works further explore the consequences of abandoning rotational invariance by constructing a basis of generalized pseudo-potentials for two-body effective interactions \cite{YangBo2017}, examining its implications with band mass anisotropy \cite{YangBo2012, YangKun2013, ZhuZheng2017}, and the recent study of an emergent universal property of FQH liquids \cite{Haldane2023}.}

As we've argued, the recent extension of this class of models actually makes the ``1D lattice'' presentation indispensable. It is for the above reasons that we advocate the view that known FQH parent Hamiltonians should be regarded as a broader class of solvable 1D models whose significance is on par with other analytically tractable models in 1D, such as 1D integrable models or models with a factorized wave functions (the latter two categories having non-trivial overlap\cite{Ogata1990, Seidel2004a, Ribeiro2006, Kruis2004}). From a ``1D point of view'', one can distinguish two approaches to these models: i) A hybrid approach where a (linear) generating set for the zero mode space is postulated/identified in first quantization, but then the completeness of the space so generated is proven using second-quantized squeezing techniques. This approach is extremely powerful in a mixed-Landau-level/parton state situation, where said completeness is otherwise hard to establish. ii) A fully second-quantized machinery.  Here, somewhat in the spirit of Ref. \onlinecite{Haldane2011}, the usual polynomial picture characteristic of FQH trial states is completely abandoned, and zero mode spaces are constructed and established entirely in second quantization.
The latter approach has so far been sparsely explored. It has been achieved for the Laughlin state\cite{Chen2015a, Mazaheri2015a, Schossler2022}, and for composite fermion states\cite{Chen2019a, Bandyopadhyay2020}. We also mention the rich Jack-polynomial literature\cite{Bernevig2008, Bernevig2008a}, which offers a powerful way to achieve a second-quantized representation of states with Jack polynomial wave functions. 
What we wish to do here is to further explore the direct connection between such a second-quantized representation, and the existence of a 1D, second-quantized parent Hamiltonian.  For this, we will emphasize the matrix product state (MPS) representation for fractional quantum Hall states. {MPS \cite{Klumper1993, Perez-Garcia2006, Orus2014}, which are one-dimensional tensor networks, are also known as tensor-train decomposition in computer science and mathematics \cite{Oseledets2009, Oseledets2009b, Oseledets2009c, Bachmayr2016}.}
To our knowledge, while there have been seminal developments in understanding FQH states
as MPS,\cite{Hansson2007a,Dubail2012a,Zaletel2012b,Tong2016,Estienne2013c,Wu2015,Crepel2018,Kjall2018} it is only for the Laughlin state that the existence of a frustration-free parent Hamiltonian has been understood as a direct consequence of the underlying MPS structure\cite{Schossler2022}. In contrast, we note that in the context of
short-ranged  models, the existence of a (finite bond dimension) MPS/tensor network \cite{Orus2014,Perez-Garcia2006} ground state is the bread-and-butter of the study of frustration-free parent Hamiltonians, such as the AKLT model\cite{Affleck1987}. 
For FQH parent Hamiltonians, this direct connection is considerably obscured by the long-ranged character of the Hamiltonian (in the 1D lattice formulation), and the related infinite bond-dimension MPS. 
We feel that a more thorough understanding of the direct relationship between the (infinite bond dimension) MPS structure of the ground state and the existence of (long-ranged) frustration-free 1D parent Hamiltonian is instrumental for further progress. For one, it is through this connection that the correspondence between edge theory and zero mode spaces becomes most manifest. More importantly, we feel that a more thorough understanding may be instrumental for further generalization.
In this paper, we will thoroughly expose this connection in the context of the non-Abelian Moore-Read (MR) fractional quantum Hall state.\cite{Moore1991} That is, we show how the MPS structure of these states defined in terms of CFT data allows for the understanding of the existence of a frustration-free three-body parent Hamiltonian. In particular, such a Hamiltonian can be established without resorting to the polynomial description of the MR state.


The remainder of this paper is organized as follows: In Section \ref{genFrameworkSection}, we develop a comprehensive framework that establishes a connection between the infinite bond dimension MPS of fractional quantum Hall states and the presence of frustration-free parent Hamiltonians for such states. Our focus is on elucidating the overarching nature of this framework and its potential applicability within a wider context.
In Sections \ref{MR-CFT_section} and \ref{sec_MR_H_and_Induct}, we apply this framework specifically to non-Abelian Moore-Read (MR) fractional quantum Hall states. In Section \ref{MR-CFT_section}, we review the Moore-Read state and its CFT as well as associated expressions for the CFT-MPS representation of this state and its zero-mode excitations. In Section \ref{sec_MR_H_and_Induct}, we present the MR parent Hamiltonian in second quantization and prove the two prerequisites required by the inductive framework introduced in Section \ref{genFrameworkSection}. This proof eliminates the necessity of any detour using first quantized-polynomial techniques
in
demonstrating that the CFT-MPS of the Moore-Read state is associated with a frustration-free Hamiltonian. We conclude with final remarks and an outlook in the Section \ref{conclusion}.

\section{General Framework\label{framework} } \label{genFrameworkSection}
{{In previous work \cite{Schossler2022}, beginnings of a framework were developed to connect the infinite
bond dimension MPS of CFT-fractional quantum Hall states to the existence of frustration-free parent Hamiltonian for the same states. However, the only state thoroughly examined in this framework has been the Laughlin state (at general $\nu=1/q$), and its quasi-hole/edge type excitations. It is thus not clear what features of the formalism were generic, and what features were intrinsic to the Laughlin state, arguably the simplest case of an FQH state.}}
In this Section, we will summarize the key ingredients used in the proof that a given class of CFT-MPS wave functions admits a frustration-free parent Hamiltonian. We will expose the generic properties that will make generalization to other classes of CFT-MPS wave function, and even more general wave function constructions, straightforward in principle. {In subsequent sections, we will rigorously apply this framework in the context of the non-Abelian Moore-Read (MR) fractional quantum Hall states. In other words, we demonstrate that by defining these states using a CFT-MPS, we can comprehend the existence of a frustration-free three-body parent Hamiltonian, without resorting to traditional first-quantized polynomial techniques.}

{To begin,} we introduce a $k$-body Hamiltonian of the form
\begin{subequations}\label{DEFgenHkbody}
\[\label{genHkbody}
H_k = \sum_r  T_r ^\dagger T_r\,,
\]

where

\[\label{genT}
T_r = \sum_{j_1\dotsc j_k} \eta^r_{j_1\dotsc j_k} c_{j_1}\dots c_{j_k}
\]
\end{subequations}
is the destruction operator associated with a $k$-particle state
${T_r}^\dagger \ket{0}$. Here, $r$ is taken from some countable index set.
Similarly, the indices $j_i$ label a basis of single-particle orbitals, and are likewise taken from a countable index set. 

One may further consider general Hamiltonians 
\[\label{genH}
H= \sum _{k\leq K} H_k
\]
with up to $K$-body interactions. 
This will be the case for the Moore-Read parent Hamiltonians discussed below, with $K=3$.
However, by positivity, imposing the zero mode condition for $H$ is equivalent to imposing the zero mode conditions for all $H_k$ jointly.
It suffices to study the zero mode condition for one $H_k$ at a time.
We thus now focus on one such $H_k$.

Let ${\cal H}_N$ be the $N$-particle subspace of the Fock space. Our task is accomplished if we show that for each $N$, 
the subspace of zero modes ${\cal H}^0_N\subset {\cal H}_N$ of the positive semi-definite Hamiltonian $H_k$ is non-trivial. In this paper, we { propose and examine} a particular strategy to achieve this.  Our results can be summarized by the following:

{\em Theorem:}
Let ${\cal W}_N\subset {\cal H}_N$ be $N$-particle subspaces. If ${\cal W}_k\subset{\cal H}^0_k$, and 
\[\label{genStep}
    c_j {\cal W}_N \subset {\cal W}_{N-1}\
\]
for all $j$ and all $N>k$, then ${\cal W}_N\subset{\cal H}^0_N$ for all $N\geq k$.

According to this theorem, we only have to establish that the $k$-particle zero mode space is a non-trivial
subspace ${\cal W}_k\neq \{0\}$, {\em and } that \Eq{genStep} holds for some likewise
non-trivial sequence of subspaces ${\cal W}_{N>k}$. 
Under the hood, the proof of this theorem is an induction proof, where the condition
${\cal W}_k\neq \{0\}$ represents the induction beginning, and \Eq{genStep}
facilitates the induction step.
To see how this works, one need only {to} observe that the following identity holds {(Appendix \ref{ident_show})}:


\[\label{genTident}
    T_r  (\hat N-k) =(-1)^{k\xi} \sum_j c^\ast_j T_r c_j\,
\]
where $\xi=1$ for fermions, $\xi=0$ for bosons, and $\hat N=\sum_j c^\ast _j c_j$ is the particle number operator.
More precisely, in writing this, we have assumed that the operators $c_j$ are pseudo-fermion or -bosons destruction operators\cite{Bagarello2017, Chen2018a}, respectively. That is, together with the creation operators $c_j^\ast$, they satisfy the familiar algebra
$c_jc_{j'}^\ast -(-1)^{\xi} c_{j'}^\ast c_j=\delta_{j,j'}$.
The only subtle difference between these pseudo-particle operators and ordinary particle operators is the fact that the Hermitian adjoint $c^\dagger_j$ of $c_j$ need not agree with $c^\ast_j$ (though there are linear relations between these two sets of operators). Physically, this corresponds to the situation where the orbitals associated with the $c_j$ are not orthogonal and/or not normalized. For this section, the reader not interested in pseudo-creation and annihilation operators may restrict attention to the special case $c_j^\dagger = c_j^\ast$.
For the following sections, it will, however, be important  {that} all arguments will work for pseudo-creation/annihilation operators as well as ordinary ones. In either case, however, we will use the dagger $\dagger$ in \Eq{genHkbody}, to ensure the Hermiticity and positive semi-definiteness of the Hamiltonian. It is also worth emphasizing that the familiar identity for the particle number operator given below \Eq{genTident} does also hold for pseudo-particle operators.

The proof of the Theorem is now a straightforward induction: Consider
the induction assumption ${\cal W}_{N-1}\subset{\cal H}^0_{N-1}$, for some $N>k$. Then, consider a
$\ket{\psi_N}\in {\cal W}_N$. 
By positive semi-definiteness, the state $\ket{\psi_N}$ is annihilated by each term $T_r ^\dagger T_r$ of $H_k$, which, moreover, is equivalent to saying that it is annihilated by each of the operators $T_r$. The zero mode condition 
$\ket{\psi_N}{\in\,} {\cal H}^0_N$, which we wish to demonstrate, can thus equivalently be stated as 
\[\label{genZMC}
T_r \ket{\psi_N} = 0 \;\;\forall\;\;r\,.
\]
We show that this follows from the assumptions about $\ket{\psi_N}$ by considering the right hand side of the last equation, and multiplying by $N-k\neq 0$:
\begin{align}
    (N-k)T_r\ket{\psi_N}&=T_r(\hat N-k)\ket{\psi_N}\nonumber\\
    &=(-1)^{k\xi} \sum_j c^\ast_j T_r c_j\ket{\psi_N}\nonumber\\
    &=0, \label{genIndStep}
\end{align}
In the second line, we utilized the identity given by equation (\ref{genTident}). The assumption $c_j\ket{\psi_N}\in{\cal W}_{N-1}\subset{\cal H}^0_{N-1}$, then implies $T_r c_j\ket{\psi_N}=0$, giving the last line.
\Eq{genZMC} then follows from $N-k\neq 0$, thus showing ${\cal W}_N\subset{\cal H}^0_N$. { The base case of the induction, ${\cal W}_k\subset{\cal H}^0_k$}, was a prerequisite of the Theorem. This completes the induction and proves the Theorem \qedsymbol{}.


In this section, we emphasize the general character of this Theorem. 
Subsequent sections will be devoted to
applications to CFT-MPS states in the mold of certain fractional quantum Hall trial wave functions, specifically the Moore-Read states. This is a natural playground for this theorem, since, as we will argue, \Eq{genStep} is a natural property of such trial wave functions, and at the same time, direct application of the Hamiltonian is quite non-trivial (in the MPS representation).


For the general relation of this approach with others found in the literature, a few additional remarks are in order.
It is clear that for $k$-body operators, $N=k$ represents the smallest $N$ suitable for an induction beginning, as the zero-mode property is trivial for $N<k$. 
{We contrast this with the case of { Ref. \onlinecite{LiAlgebraicMR}}, where a scheme was employed that is similar but more tailored to the case where incompressible ground states satisfy a certain recursion relation (note that this induction assumption in that reference is about the incompressible state only, not the full zero mode sector, and the induction step proceeds by adding {\em two} particles). There, while $k=3$, an induction beginning of $N=6$ was found necessary. We expect the scheme introduced here to be considerably more general. 

Indeed, we expect {this} scheme to  apply to a large number of CFT-MPS states formulated in the quantum Hall context. In particular, we believe the property \eqref{genStep} to be a generic property of such variational states. 
On physical grounds, such states are expected to yield a complete set of incompressible states {\em and} quasi-hole like excitations, and the removal of one electron (or bosonic particle) can generically be understood as the introduction of a certain cluster of quasi-holes. Therefore, the right-hand side of \Eq{genStep} cannot lead outside the zero mode space, and the equation follows, so long as the ${\cal W}_N$ represent a complete description of such zero modes.
Thus, \Eq{genStep} should be considered a requirement for a good variational description in this context, and the task of finding a parent Hamiltonian then boils down to finding a Hamiltonian of the generic form \eqref{genH} for which induction beginnings can be established. There is nothing, however, in this scheme that is limited to lowest Landau (symmetric polynomial) wave function with ``nice clustering properties''. Thus, we expect that this scheme may prove fruitful
beyond the lowest Landau level, beyond the quantum Hall context, and beyond the context of formally one-dimensional Hamiltonians.

\section{Moore-Read CFT}\label{MR-CFT_section}


{The framework detailed above generalizes the one 
introduced in Ref. \onlinecite{Schossler2022} to study the connection between Laughlin state parent Hamiltonians and the MPS representation
of their ground states.
The latter are arguably the simplest in the CFT-MPS class.}
To demonstrate the generalizability of this scheme along the lines of the preceding section, we consider now the Moore-Read state as a concrete example.
That is, we wish to understand the zero-mode property of this non-Abelian quantum Hall state and its
bosonic as well as Majorana-like 
excited states strictly from an MPS point of view, given the state's frustration free parent Hamiltonian.
 The latter has, of course, originally been obtained from the first-quantized, polynomial representation of these states.\cite{Greiter1992} In this work, we will only use this polynomial representation to review its relation to the MPS, but will otherwise not use it.
 
 {

To start, let us review the Moore-Read state and its connection to CFT correlators. The Moore-Read state at filling factor $\nu=1/q$,
omitting Gaussian factors, is given by the following polynomial: \cite{Moore1991,Nayak1996,Nayak2008,Hansson2016}
\begin{align}
\psi\left(z_{1},\cdots,z_{N}\right) & =\left\langle \psi_{e}\left(z_{N}\right)\cdots\psi_{e}\left(z_{1}\right)\right\rangle \\
 & =\text{Pf}\left(\frac{1}{z_{i}-z_{j}}\right)\prod_{i<j}\left(z_{i}-z_{j}\right)^{q}\,,\label{first-quant-moore-read}
\end{align}
where the particle operator for this state is the product of a chiral Majorana
field in the Ising CFT and the vertex operator of a free massless chiral boson CFT: $\psi\left(z\right)=\chi\left(z\right)\times V\left(z\right)$.
The bosonic or Coulomb sector is analogous to similar presentations for the Laughlin state. We proceed by summarizing some of the most important properties of this CFT, and refer the interested reader to other references \cite{di1996conformal,mussardo2009statistical}\cite{Dubail2012a,Zaletel2012b,Estienne2013c,Crepel2018}. The holomorphic (chiral) part of the vertex operator,  $V_{\sqrt{q}}\left(z\right)=:e^{i\sqrt{q}\,\phi\left(z\right)}:$, generates the 
Jastrow factor\cite{FUBINI1991,Cristofano1991,Moore1991,di1996conformal,mussardo2009statistical, Zaletel2012b,Tong2016,Estienne2013c,Wu2015,Crepel2018,Kjall2018} in (\ref{first-quant-moore-read}),
\begin{equation}
\left\langle V_{\sqrt{q}}\left(z_{N}\right)\cdots V_{\sqrt{q}}\left(z_{1}\right)\right\rangle. \label{statate_correlator}
\end{equation}
More precisely, $V_{\sqrt{q}}\left(z\right)$ is
a primary field in a chiral-free massless bosonic  CFT in $1+1d$ with
$U(1)$ charge $\sqrt{q}$. It can be given the mode expansion  
\begin{equation}
V_{\sqrt{q}}\left(z\right)=\sum_{\lambda}V_{-\lambda-h}z^{\lambda},\label{vertex_mode_exp}
\end{equation}
where $h=q/2$ is the conformal dimension of $V_{\sqrt{q}}$.  The neutral excitations of this theory can be expressed in terms of modes $a_n$, which are the modes of the chiral bosonic field,
\begin{equation}
\phi\left(z\right)=\phi_{0}-ia_{0}\log\left(z\right)+i\sum_{n\neq0}\frac{1}{n}a_{n}z^{-n}.\label{phimode}
\end{equation}
The $a_{n}\text{'s}$ obey the algebra: $\left[\phi_{0},a_{0}\right] =i$ and $\left[a_{n},a_{m}\right] =n\delta_{n+m,0}$. Additionally, their action on states can be described as follows:
\begin{align}
a_{n}\ket{N} & =0,\quad n>0\nonumber \\
\bra{N}a_{n} & =0,\quad n<0\label{a_nPsi} \\
a_{0}\ket{N} &=\sqrt{q}N\ket{N},\label{a_0psiN}
\end{align}
where $\ket{N}$ is a primary state of the bosonic CFT.

In the Ising sector of the CFT, the Majorana field
$\chi\left(z\right)$ can be chosen to have one of two monodromy properties
\cite{di1996conformal,mussardo2009statistical}. With periodic
or Neveu-Schwarz (NS) boundary condition, the correlator of two $\chi\left(z\right)$
is given by
\[
\left\langle \chi\left(z\right)\chi\left(w\right)\right\rangle =\frac{1}{z-w},
\]
so the correlator of $N$ of these fields generates the Pfaffian
\[
\left\langle \text{out}|\chi\left(z_{1}\right)\cdots\chi\left(z_{N}\right)|\text{0}\right\rangle =\text{Pf}\left(\frac{1}{z_{i}-z_{j}}\right).
\]
Here, the $\bra{\text{out}}$-state 
must be defined with care depending on whether the particle number is odd or even, where in the former case, it must contain a Majorana mode. We give details below.
In this NS sector, on the
plane, one has the mode expansion of the chiral Majorana field
\begin{align}
\chi\left(z\right) & =\sum_{n\in\mathbb{Z}}\chi_{n-1/2}z^{-n}\,,\label{majorana_mode_exp}
\end{align}
which implies 
\begin{equation}
\chi_{n-1/2}=\ointop\frac{dw}{2\pi i}w^{n-1}\chi\left(w\right),\quad n\in\mathbb{Z}.\label{xi_mode_expansion}
\end{equation}
The positive (negative) index modes annihilate the vacuum when acting
from the left (right): 
\begin{align}
\chi_{n-1/2}\ket{0} & =0,\quad n>0\nonumber \\
\bra{0}\chi_{n-1/2} & =0,\quad n\leq0.\label{conditions_modes_majo}
\end{align}
These Majorana modes obey the anticommutation algebra
\begin{equation}
    \left\{ \chi_{n-1/2},\chi_{m-1/2}\right\} =\delta_{n+m,1},\ n,m\in\mathbb{Z}.\label{majo-anticommut}
\end{equation}
The correlator of two Majorana fields in the Ramond sector (or antiperiodic
sector) does not generate the Pfaffian function, so we will work only
with the NS sector of this free (Majorana) fermion theory. The twist
operator of the Ising CFT connects these two sectors.

Moreover, the twist operator is used for the creation of localized quasi-holes when inserted into the correlator.
However, quasi-holes are always created in pairs,\cite{Moore1991}
and a pair of twist operators can fuse to the identity or to a $\chi$ only.
Therefore, to generate a complete set of edge excitations, hence zero modes,
it should be sufficient to insert 
operators $\chi_{r-1/2}$ into the Ising part of the 
correlator,\cite{Dubail2012a} as in the expression 
\begin{equation}
\left\langle 0| \chi_{r_{1}-1/2}\cdots\chi_{r_{F}-1/2}\chi\left(z_{N}\right)\cdots\chi\left(z_{1}\right)|0\right\rangle\, , \label{Majorana-sector}
\end{equation}
apart from  bosonic excitations that will similarly correspond to insertions of the $a_n$, see below.
In Eq. \eqref{Majorana-sector},
 the condition $\left(N-F\right)\in2\times\mathbb{N}$ must also be 
satisfied to ensure that the expression, which still can be written as a Pfaffian, is non-zero. 
This implies that excitations $\chi_{r-1/2}$
must be added in pairs, except for a single extra
factor that we may choose to be $\chi_{1/2}$ when $N$ is odd (see below). Also, the restriction
to $N-F$ even 
represents a superselection rule
between the charged (Coulombic) sector and the Majorana sector 
that is somewhat reminiscent 
of global
selection rules governing spin-charge separation in the edge theory of Halperin
states\cite{Milovanovic1996}. The anti-commutation relation of $\chi_{r-1/2}$ implies that in Eq. \eqref{Majorana-sector}, these modes have to be distinct from one another, except for the extra $\chi_{1/2}$ in the odd case, which we will, however, absorb into the $\langle\text{out}|$-bra below.
Also, we must have $r_{i}\geq1$ for non-zero result in  Eq. \eqref{Majorana-sector}, as stated by \eqref{conditions_modes_majo}.



The CFT-MPS representation of the Moore-Read state can be obtained by combining the Majorana field and the vertex operator mode expansion in eqs.  (\ref{majorana_mode_exp}) and (\ref{vertex_mode_exp}), respectively.
For details, we refer the reader to the pertinent literature.\cite{Dubail2012a,Zaletel2012b,Tong2016,Estienne2013c,Wu2015,Crepel2018,Kjall2018}
This yields the following expression:
\begin{align}
\psi\left(z_{1},\cdots,z_{N}\right)= & \sum_{\left\{ k_{i}\right\} }\sum_{\left\{ \lambda_{i}\right\} }\left\langle \chi_{\lambda_{N}-k_{N}-1/2}\cdots\chi_{\lambda_{1}-k_{1}-1/2}\right\rangle \nonumber \\
 & \times\braket{N|V_{-\lambda_{N}-h}\cdots V_{-\lambda_{1}-h}|0}\prod_{i=1}^{N}z_{i}^{k_{i}}\label{MR-mode-expansion}
\end{align}
where $\left\{ k_{i}\right\}$, $\left\{ \lambda_{i}\right\}$ denote unrestricted sets of index-variables. We can introduce an ordered set of index-variables via
\begin{equation}
\left(k_{i}\right)^{N}:\begin{cases}
k_{N} & \geq\cdots\geq k_{1}\text{ for bosons}\\
k_{N} & >\cdots>k_{1}\text{ for fermions},
\end{cases}\label{def_k_{i})^N}
\end{equation}
and translate the Moore-Read state in (\ref{MR-mode-expansion}) to a second-quantized language by noticing
that for fixed $\left\{ k_{i}\right\}$, the sum over $\left\{ \lambda_{i}\right\}$ renders the product of correlators (anti-)symmetric in the $k_i$ for $q$ odd (even). Thus, the sum over $\left\{ k_{i}\right\}$ (anti-)symmetrizes the product over $z_i^{k_i}$, yielding a bosonic (fermionic) occupation number eigenstate $\ket{\left(k_{i}\right)^{N}}$ with the $k_i$-orbitals occupied:
\begin{equation}
\ket{\psi_{N}}=\sum_{\left(k_{i}\right)^{N}}C_{\left(k_{i}\right)^{N}}\ket{\left(k_{i}\right)^{N}},\label{MR_MPS}
\end{equation}
where 
\begin{align}
C_{\left(k_{i}\right)^{N}} & =\frac{1}{\prod_{i}l_{i}!}\sum_{\left\{ \lambda_{i}\right\} }\left\langle \chi_{\lambda_{N}-k_{N}-1/2}\cdots\chi_{\lambda_{1}-k_{1}-1/2}\right\rangle \nonumber \\
 & \qquad\times\braket{N|V_{-\lambda_{N}-h}\cdots V_{-\lambda_{1}-h}|0},\label{coef_MR}
\end{align}
Here, $l_{i}$ is the occupancy of the $i$-th angular
momentum state, which is only relevant for bosons. For fermions $l_{i}\in\{0,1\}$.
The orbital basis elements $\ket{\left(k_{i}\right)^{N}}$ are defined via 
\begin{equation}
\braket{z_{1},\!\cdots\!,z_{N}|\left(k_{i}\right)^{N}}\!=\!\frac{1}{N!}\sum_{\sigma\in S_{N}}\left(\text{ sgn}\,\sigma\right)^{q}\prod_{i=1}^{N}z_{i}^{k_{\sigma_{i}}}.\label{slater}
\end{equation}
This implies that $\ket{\left(k_{i}\right)^{N}}=\left(\sqrt{N!}\right)^{-1}c_{k_{1}}^\ast...c_{k_{N}}^\ast\ket{0}$,
 where $c_{k}^\ast$ is a pseudo particle creation operator, which creates a particle  in a state with un-normalized
wave function $z^{k}\exp\left(-|z|^{2}/4\right)$. As mentioned in Sec. \ref{framework}, the associated destruction operators $c_k$ may be defined such that $[ c_k, c_{k'}^\ast]_{\pm}=\delta_{k,k'}$ holds. We emphasize again that $c_k^\ast$ is {\em not} the Hermitian adjoint of $c_k$, however, simple re-scaling $c_k\rightarrow {\mathcal N}_k c_k$, $c_k^\ast\rightarrow {\mathcal N}_k^{-1} c_k^\ast$ does turn these pseudo creation/annihilation operators into ordinary ones. In particular, the particle number operator is still $\hat N=\sum_k c^\ast_k c_k$.

A more concise notation for these coefficients can be achieved by expressing them in terms of the modes of the particle-field  operator:
\begin{equation}
    C_{\left(k_{i}\right)^{N}}=\frac{1}{\prod_{i}l_{i}!}\braket{\text{out}|\psi_{-k_{N}-h-1/2}\cdots\psi_{-k_{1}-h-1/2}|\text{in}},\label{coef_densest_MR}
\end{equation}
where 
\[
\psi_{-k-h-1/2}=\sum_{\lambda}\chi_{\lambda-k-1/2}V_{-\lambda-h}.\label{psi-mode:chiV}
\]
The modes of this particle operator satisfy (anti-)commutation relations for fermions and bosons, respectively.
The relevant range of the $k_{i}\text{'s}$
as well as their total angular momentum
$\sum_i k_i$ are governed by conservation laws in the CFT as well as the choice of in- and out-states, $\ket{\text{in}}$ and $\ket{\text{out}}$.
For the
Moore-Read state, the proper choice, 
which was implicitly assumed already in Eq. {\eqref{first-quant-moore-read}}, is 
\begin{equation}
    \bra{\text{out}}=\bra{\text{out}}_{\chi}\otimes\bra{\text{out}}_{V}
\end{equation}
where the states $\bra{\text{out}}_{\chi}$ and $\bra{\text{out}}_{V}$ are associated with the Majorana and Coulomb gas CFT sectors, respectively. Formally,
\begin{align}
 \bra{\text{out}}_{V}\doteq\bra{N},
\end{align}
and
\begin{align}\label{outchi}
\bra{\text{out}}_{\chi}\doteq\begin{cases}
\bra{0}\!\! & \text{if }N\text{ even,}\\
\bra{0}\chi_{1/2}\!\! & \text{if }N\text{ odd}.
\end{cases}
\end{align}
The mode $\chi_{1/2}$ is necessary for describing a system with an odd number of particles, and it corresponds to a Majorana fermion that is located at infinity.\cite{Dubail2012a} The $\ket{\text{in}}$ state is defined as the respective vacuum in both CFT sectors,
\begin{equation}
\ket{\text{in}} =\ket{0}\otimes\ket{0}.
\end{equation}
The bra $\bra{N}$ can be interpreted as the result of the background charge operator acting on the Coulomb gas CFT vacuum bra, ensuring charge neutrality. In the following, we will have
the need to evaluate expressions where $\bra{N}$ is acted upon on the right by $V_{-\lambda-h}$. This motivates the introduction of an alternative set of neural bosonic modes that generate the same algebra as the $a_{n>0}$. Indeed, $V_{-\lambda-h}$ {\em removes} one unit of charge from $\bra{N}$, so one may write
\begin{align}
\bra{N}V_{-\lambda-h} =\bra{N-1}\sum_{l=0}^{q\left(N-1\right)-\lambda}\!\!\!\!\!b_{q\left(N-1\right)-\lambda}^{l},\label{bra_N_V_lambda}
\end{align}
where $b_{k}^{l}$ are a collection of $l$ neutral excitations 
related to the $a_{n>0}$ via\cite{Schossler2022}
\begin{align}
b_{k}^{l} & =\frac{\left(-\sqrt{q}\right)^{l}}{l!}\!\!\!\!\!\!\sum_{i_{1}+\cdots+i_{l}=k}\!\!\frac{a_{i_{1}}}{i_{1}}\frac{a_{i_{2}}}{i_{2}}\cdots\frac{a_{i_{l}}}{i_{l}},\, i_{j}>0\Rightarrow k\geq l;\nonumber \\
b_{k}^{0} & =\begin{cases}
1,\! & k=0\\
0,\! & k>0.
\end{cases}\label{b_k^l_def-1}
\end{align}
Thus far, the MPS \eqref{MR_MPS} generates the incompressble MR-state, both for even and odd particle number. To generate a complete set of zero modes for the frustration free MR-parent Hamiltonian to be discussed below, we need to introduce additional bosonic and fermionic mode operators modifying the $\bra{\text{out}}$ bra, as anticipated earlier. 
Formally, a basis of zero mode states can be obtained in MPS form via the following Ansatz:
\begin{align}
    \ket{\psi_{N}^{a_{n}\cdots\chi_{r}\cdots}} =  \sum_{\left(k_{i}\right)^{N}}C^{{n}\cdots{r}\cdots}_{\left(k_{i}\right)^{N}}\ket{\left(k_{i}\right)^{N}},\label{MR_MPS_qh_basis}
\end{align}
where the MPS coefficients have additional superscripts ${n}\cdots{r}\cdots$ to indicate the excitations that have been added in the bosonic sector, $n...$, and in the Majorana sector, $r...$:
\begin{align}
    C^{{n}\cdots{r}\cdots}_{\left(k_{i}\right)^{N}} = \frac{1}{\prod_{i}l_{i}!}\bigl\langle\text{out}|a_{n}\cdots\chi_{r}&\cdots\psi_{-k_{N}-h-1/2}\cdots\nonumber \\
    \cdots&\psi_{-k_{1}-h-1/2}|\text{in}\bigr\rangle.\label{coef_excit_MR}
\end{align}
Here, $a_{n}\cdots\chi_{r}\cdots$ denotes and finite string of $a_n$ and (even number of) $\chi_r$ operators, e.g., $a_{1}a_{2}a_4\chi_{3/2}\chi_{5/2}$.
If the associated superscripts are omitted, i.e., the string is empty, we are referring  to the coefficients defined in Eq. (\ref{coef_MR}) and (\ref{coef_densest_MR}), which yield the incompressible MR state. In the following,
$a_{n}\cdots\chi_{r}\cdots$ will always represent a string that may or may not be empty 
in both the bosonic sector and the Majorana sector, i.e., in particular, the notation  \eqref{MR_MPS_qh_basis} may or may not refer to the incompressible state unless further specified.
As both the $a_{n}$- as well as $\chi_{r}$-insertions increase the angular momentum of the state by an amount equal to the respective subscripts, the incompressible MR will be the ``densest'' zero mode in the sense of lowest angular momentum.
\section{Moore-Read Hamiltonian and the zero mode property induction} \label{sec_MR_H_and_Induct}


As {it} is well established,
the Moore-Read state is the densest zero mode of a local 3-body parent Hamiltonian.\cite{Greiter1991,Greiter1992,Greiter1992a, Read1996, Simon2007a}
For $q=1$ and $q=2$, the latter is easily expressed in terms of delta-functions or derivatives thereof. Second quantized expressions for these cases have been given in the literature.\cite{Seidel2006,Bergholtz2006, Weerasinghe2014a}
In general, the 3-body operator must give positive energy to any three particles in a state of relative angular momentum $3q-2$ or less. Alternatively,
it must give positive energy to any three particles in a state of relative angular momentum equal to $3q-3$ if additional 2-body operators are present\cite{Read1996}.
There is no loss of generality in discussing the former variety. We will discuss the relevant 2-body operators later below.
A second-quantized Hamiltonian giving finite energy to 3-body states with relative angular momentum $L_{\text{rel}}\leq 3q-2$ can be constructed as follows:
\begin{subequations}\label{H_3bd}
\begin{equation}
H_{\frac{1}{q}}^{\text{3bd}}=\sum_{\substack{0\leq |t|<(3q-1)}
}\sum_{J\geq 0}T_{J}^{t\dagger}T_{J}^{t},
\end{equation}
where 
\begin{equation}
T_{J}^{t}=\sum_{m+n+p=J}f^{t}\left(m,n,p\right)c_{m}c_{n}c_{p}.\label{def_T}
\end{equation}
\end{subequations}
The operator $T_{J}^{t}$ annihilates three particles in a state with a total angular momentum of $J$, labeled by a multi-index $t=(t_1, t_2, t_3)$, and 
$f^t$ is a polynomial in $m,n,p$ of degree $|t|=t_1+t_2+t_3$ (except in those cases where it vanishes) defined by
$f^{t}={\mathcal S}\,\left(m-n\right)^{t_1}\left(m-p\right)^{t_2}\left(n-p\right)^{t_3}$.
Here, $\mathcal S$ denotes the (anti-)symmetrizer in $m,n,p$ for $q$ odd (even). In the sum in \Eq{H_3bd}, it is also implied that all $t_i\geq 0$.
Despite appearances, with the form factor being polynomial in the separation between orbitals,
this interaction is exponentially cut off at very large distances. To see this, we pass to ordinary creation/annihilation operators, 
which brings back the aforementioned normalization factors ${\mathcal N}_k$.
Clearly $H_{{1/q}}^{\text{3bd}}$ is positive (semi-definite), and moreover
is of the general form \eqref{DEFgenHkbody} with $k=3$ and the index $r$ corresponding to the multi-index $(t,J)$. We re-state the zero mode condition \eqref{genZMC} here for this special case as
\[
     {T_J^t\ket{\psi_{N}^{a_{n}\cdots\chi_{r}\cdots}} =0,}\label{T-ind-assump_3}\\
\]
for all $t$ and $J$ that appear in the sum \eqref{H_3bd}. To see that this second quantized Hamiltonian has all the desired properties, we first observe that
and Hamiltonian of the form \eqref{H_3bd} should be regarded as just one representative of an entire {\em class} of Hamiltonians with identical zero mode space. Indeed, replacing the $T^t_J$ with new linearly independent linear combinations of themselves (in particular, at fixed $J$, which is of greatest interest) leads to equivalent zero mode conditions \eqref{T-ind-assump_3}. Now, a 3-particle state with total angular momentum $J$ and relative angular momentum $L_r$ can be created by an operator whose adjoint is of the general form \eqref{def_T}, i.e., with some form factor $f(m,n,p)$. The latter will be a polynomial of degree $L_r$. Now, any such $f(m,n,p)$ can be linearly generated from terms of the form  $\left(m-n\right)^{t_1}\left(m-p\right)^{t_2}\left(n-p\right)^{t_3}(m+n+p)^{t_4}$, with $L_r=\sum_{i=1}^4 t_i$. In such expressions, however, the term $(m+n+p)^{t_4}$ is just a constant when plugged into \Eq{def_T}. It follows thus that the zero mode condition associated to any 3-particle state of relative angular momentum less than $3q-1$ can be obtained from \Eq{T-ind-assump_3}. 
One might still worry if by summing over all possible $t$, we have kept ``too many'' zero mode conditions. This, however, turns out not to be the case, since we will show below that Eq. \eqref{H_3bd} has all the zero modes any
``Moore-Read parent Hamiltonian'' is supposed to have.

As mentioned, the MR-state and its hole-like excitations/edge excitations are also annihilated by certain two-body operators, which we discuss next.
Let
\begin{subequations}\label{H_2bd}
\begin{equation}
H_{\frac{1}{q}}^{\text{2bd}}=\sum_{\substack{0\leq t<q-1\\
\left(-1\right)^{t}=\left(-1\right)^{q-1}
}
}\sum_{J}Q_{J}^{t\dagger}Q_{J}^{t},
\end{equation}
where 
\begin{align}
Q_{J}^{t} & =\sum_{x=-J/2}^{J/2}x^{t}c_{J/2-x}c_{J/2+x},\label{def_Q}
\end{align}
\end{subequations}
and $t$ now labels an ordinary non-negative integer. 
For $t=0,1$, one obtains the Haldane pseudo-potential projections onto relative angular momenta $0$ (bosons) and $1$ (fermions), respectively.
Again, this becomes manifest by passing from pseudo creation/annihilation operators to ordinary ones. For $t>1$, the relation with Haldane pseudo-potentials becomes more complicated, however, the zero mode condition derived from Eq. \eqref{H_2bd},
\[
{Q_{J}^{t}\ket{\psi_{N}^{a_{n}\cdots\chi_{r}\cdots}} =0,}\label{Q-ind-assump_3}
\]
for all $t$ and $J$ appearing in Eq. \eqref{H_2bd},
is exactly equivalent to that obtained by summing the Haldane pseudo-potential projections with indices less than $q-1$ (keeping only oven or odd ones as in Eq. \eqref{H_2bd}). 
This is so since in order to pass to bona fide Haldane pseudo-potentials, we only need to form certain linearly independent new linear combinations of the $Q^t_J$, and reason in the same way as done above for the 3-body case. 

{Although this section primarily considers the disk geometry, the same principles are applicable to other genus-zero geometries, such as the cylinder and sphere. In these geometries, the MR state remains consistent, characterized by appropriately defined pseudofermions\cite{Ortiz2013}.
Going to genus-one geometries, such as the torus geometry, is not straightforward but is worth exploring in the future. The MPS representation of several FQH states has also been explored in various geometries\cite{Estienne2013c}, as well as for the Haldane-Rezayi state in the torus geometry.\cite{Crepel2019}}

We note now that the combined Hamiltonian
\[\label{Hfull}
H_{\frac{1}{q}}= H_{\frac{1}{q}}^{\text{2bd}}+H_{\frac{1}{q}}^{\text{3bd}},
\]
which serves as a parent Hamiltonian for the Moore-Read state at filling factor $1/q$, is a $K=3$ special case of the general Hamiltonian discussed in Sec. \ref{framework}, with $k=2$ and $k=3$ terms present. The zero mode condition
of $H_{\frac{1}{q}}$ is the combined zero mode condition associated with $H_{\frac{1}{q}}^{\text{2bd}}$ and with $H_{\frac{1}{q}}^{\text{3bd}}$. We may thus study its zero modes by connecting with the Theorem of Sec. \ref{framework},
where we identify the spaces ${\cal W}_N$ with the spaces spanned by the MPS states $\ket{\psi_{N}^{a_{n}\cdots\chi_{r}\cdots}}$ (for given $N$).
To show that these states are zero modes, all we need to do is hence to ensure that the prerequisites of the Theorem are met.
That is, we demonstrate the induction step \eqref{genStep}, and the induction beginnings for these classes of MPS.
Only the induction beginnings must be done separately for $H_{\frac{1}{q}}^{\text{2bd}}$ and for $H_{\frac{1}{q}}^{\text{3bd}}$. The inductions step \eqref{genStep} is universal.
We thus begin with the latter.

\subsection{The induction step} \label{MRind}

{From the above, we} need to prove
\Eq{genStep} for the specific situation at hand.
Note that for this step, we will not require detailed knowledge of the operators $T^t_J$ and $Q^t_J$. Instead, the zero mode property will apply to any Hamiltonian of the generic makeup of Eqs. \eqref{H_3bd} and \eqref{H_2bd} (including $k$-body generalizations) for which the induction beginning 
${\cal W}_k \subset H^0_k$
can be proven. Hence, we defer the latter and turn to the induction step, $c_j {\cal W}_N \subset {\cal W}_{N-1}$.
The crucial ingredient  is thus to analyze
the action of an annihilation operator $c_k$ on the MPS state \eqref{MR_MPS_qh_basis}.
{Therefore, we} investigate
\begin{align}
\sqrt{N}c_{k} &\ket{\psi_{N}^{a_{n}\cdots\chi_{r}\cdots}} \!=\!\!\!\!\sum_{\left(k_{i}\right)^{N-1}}\!\!\bra{\text{out}}{a_{n}\cdots\chi_{r}\cdots}\,\psi_{-k-h-1/2}\nonumber \\
 &\quad \times\overset{N-1}{\overbrace{\psi_{-k_{N}-h-1/2}\cdots\psi_{-k_{1}-h-1/2}}}\ket{\text{in}}\ket{\left(k_{i}\right)^{N-1}},\label{c_kr_psi_N}
\end{align}
where we moved the particle mode $\psi_{-k-h-1/2}$ with index $k$ appearing in the MPS representation of $\ket{\psi_{N}^{a_{n},\chi_{r_{1}},\chi_{r_{2}}}}$, Eq. \eqref{MR_MPS_qh_basis}, all the way to the left.
This is possible because the particle operator modes $\psi_{-k-h-1/2}$ (anti-)commute for $q$ odd (even).
(Moving $\psi_{-k-h-1/2}$ then compensates a minus sign possibly arising from the action of $c_k$.)
Then, using the following (anti-) commutation relations,
\begin{align}
\left[a_{n},\psi_{-k-h-1/2}\right] & =\sqrt{q}\sum_{\lambda}\chi_{\lambda-k-1/2}V_{-\lambda+n-h}\nonumber \\
 & =\sqrt{q}\psi_{-\left(k-n\right)-h-1/2},\label{an_commut_psi}\\
 \left[a_{n},V_{-\lambda-h}\right]& =\sqrt{q}\,V_{-\lambda+n-h},\label{comut_a_n,V_lambda}\\
\left\{ \chi_{l-1/2},\psi_{-k-h-1/2}\right\}  & =\sum_{\lambda}\left\{ \chi_{l-1/2},\chi_{\lambda-k-1/2}\right\} V_{-\lambda-h}\nonumber \\
 & =V_{-\left(k-l+1\right)-h},\label{chi_commut_psi} \\ 
 \left[V_{-\lambda-h},V_{-\mu-h}\right]_{q}&
 =\left[V_{-\lambda-h},\psi_{-k-h-1/2}\right]_{q}=0 \label{v_commut_psi}
 \end{align}
 where $[\mathcal{O}_1,\mathcal{O}_2]_q=\mathcal{O}_1\mathcal{O}_2+(-1)^q\mathcal{O}_2\mathcal{O}_1$,
 we can pull the electron operator mode $\psi_{-k-h-1/2}$ even further to the left. The commutator $[\chi_{l-1/2},V_{-\lambda-h}]=[\chi_{l-1/2},a_n]=0$ because $\chi_{l-1/2}$ and $V_{-\lambda-h}$ {(or $a_n$)} are entities living in different CFT sectors.
When utilizing the relations (\ref{an_commut_psi}-\ref{chi_commut_psi}), 
it is clear that new terms appear, where either a $\psi$-mode or a $V$-mode is either acting directly on $\bra{\text{out}}$ on the right, or is still separated from $\bra{\text{out}}$ by a string of modes $a_n$ and/or $\chi_{r-1/2}$. However, with each application of the commutators, the string of modes $a_n$ and/or $\chi_{r-1/2}$ separating the $\psi$-mode or $V$-mode from $\bra{\text{out}}$ will get shorter. Hence, eventually, we will have only terms left where 
a $\psi_{-k-h-1/2}$ or a $V_{-\lambda-h}$ is acting directly on $\bra{\text{out}}$ on the right, followed by a (possibily empty) string of modes $a_n$ and/or $\chi_{r-1/2}$, followed by the string
$\psi_{-k_{N}-h-1/2}\cdots\psi_{-k_{1}-h-1/2}$
present in Eq. \eqref{c_kr_psi_N}.
We should thus evaluate $\bra{\text{out}} \psi_{-k-h-1/2}$ and $\bra{\text{out}} V_{-\lambda-h}$. It turns {out} in treating the former we will automatically cover the latter. Hence we consider

\begin{align}
 & \bra{\text{out}}\psi_{-k-h-1/2}=\bra{\text{out}}\!\!\!\!\!\!\sum_{k<\lambda\leq q\left(N-1\right)}\!\!\!\!\!\chi_{\lambda-k-1/2}V_{-\lambda-h}.\label{outpsi}
\end{align}
The modes $\chi_{\lambda-k-1/2}$ and $V_{-\lambda-h}$ operate on distinct CFT sectors. Therefore, we can split the expression in the following manner:
\begin{align}
 & \bra{\text{out}}\psi_{-k-h-1/2}=\!\!\!\!\!\!\!\!\!\sum_{k<\lambda\leq q\left(N-1\right)}\!\!\!\!\!\!\!\!\bra{\text{out}}_{\chi}\chi_{\lambda-k-1/2}\otimes\bra{\text{out}}_{V}V_{-\lambda-h}\,.
\end{align}
Now, applying $V_{-\lambda-h}$ to the $N$-particle state $\bra{\text{out}}_{V}=\bra{N}$ yields a $(N-1)$-particle $\bra{\text{out}}_{V}=\bra{N-1}$ state along with a collection of bosonic excitations $a_n$ with $n>1$, as exposed in Eq. \ref{bra_N_V_lambda}.
The net result is that Eq. \eqref{outpsi} 
can be written as 
\begin{align}
\bra{\text{out}}_{\chi}\otimes\bra{N-1}\sum_{\lambda=k+1}^{q\left(N-1\right)}\sum_{l=0}^{q\left(N-1\right)-\lambda}\!\!\!\!\!\chi_{\lambda-k-1/2}b_{q\left(N-1\right)-\lambda}^{l}\,.\label{(N-1)particles_excit}
\end{align}
The case $\bra{\text{out}} V_{-\lambda-h}$ is similar, but without the $\chi$-modes, and without the sum over $\lambda$.
As the modes $b_{q\left(N-1\right)-\lambda}^{l}$ are expressible in terms of the $a_n$ via Eq. \eqref{b_k^l_def-1}, in the end we are left with a string
${a'}_n\dots{\chi'}_r$ times $\psi_{-k_{N}-h-1/2}\cdots\psi_{-k_{1}-h-1/2}$
between the $\ket{\text{in}}$ and the $(N-1)$-particle $\bra{\text{out}}$ (this requires pulling a $\chi_{1/2}$ out of the original $N$-particle $\bra{\text{out}}_\chi$, Eq. \eqref{outchi}{, if $N$ is odd.})
Putting things together, we infer that 
Eq. \eqref{c_kr_psi_N} yields a superposition of terms of the form
$\ket{\psi_{N-1}^{{a'}_{n}\cdots{\chi'}_{r}\cdots}}$, that is

\begin{equation}
\sqrt{N}c_{k}\ket{\psi_{N}^{a_{n}\cdots\chi_{r}\cdots}}=\sum_{{a'}_{n}\cdots{\chi'}_{r}\cdots}\ket{\psi_{N-1}^{{a'}_{n}\cdots{\chi'}_{r}\cdots}}\,,
\end{equation}
where the sum goes over all strings that are generated in the process described above. {
As by definition, the stated
$\ket{\psi_{N}^{a_{n}\cdots\chi_{r}\cdots}}$ span the spaces ${\cal W}_N$, this proves
Eq. \ref{genStep} for the case under consideration, and we are done with the induction step.} 



\subsection{The induction beginning: three-particle zero mode property}
\label{indBeginning}
The Theorem in Sec. \ref{framework} relies on another prerequsite, the condition ${\cal W}_k \subset H^0_k$. Technically, this serves as the induction beginning. Interestingly, it is the only aspect of the theorem that explicitly depends on the Hamiltonian. As we are, in principle, applying 
the Theorem separately to $H_{\frac{1}{q}}^{\text{2bd}}$ and
$H_{\frac{1}{q}}^{\text{3bd}}$,
we must establish two separate induction beginnings.
We begin with the (more challenging) case of three-particle Hamiltonina $H_{\frac{1}{q}}^{\text{3bd}}$. In the present sub-section, thus, 
$H^0_N$ refers to the zero mode spaces of $H_{\frac{1}{q}}^{\text{3bd}}$.
The condition  ${\cal W}_k \subset H^0_k$ is the equivalent to Eq. (\ref{T-ind-assump_3}) for $N=3$, that is, for all states  $\ket{\psi_{3}^{a_{n}\cdots\chi_{r}\cdots}}$,
with, in principle, an arbitrary number of bosonic and/or Majorana excitations present.
We may, however, in general limit the number of each type of excitations, bosonic or Marorana, to be no more than the number of particles $N$:
A MPS with $N$ particles and more than $N$ edge excitations in either Majorana or bosonic sector can always be expressed as a linear combination of $N$-particle MPS with no more than $N$ excitations of each kind. The proof of this statement for the bosonic CFT sector is the same as the one presented for the edge excitations in the Laughlin states in Ref. \onlinecite{Schossler2022}. The proof for the Majorana sector is provided in Appendix \ref{Majo_wick_corr}.
Hence, any three-particle Moore-Read state with more than three excitations in each CFT sector can be expressed as a linear combination of states with no more than three excitations in each sector. 
Nonetheless, we still need to demonstrate the validity of the zero mode condition for the cases where $\leq N$ excitations are present in each sector, for $N=3$ and $N=2$, according to Eqs. (\ref{T-ind-assump_3}) and 
(\ref{Q-ind-assump_3}), respectively. 
 For $N=3$, there are three different scenarios:
one Majorana mode; one Majorana mode, including $\chi_{1/2}$, and one, two, or three bosonic modes; and three Majorana modes with one, two, or three bosonic modes.
Similarly, for the two-body Hamiltonian, we need to prove that the zero mode condition holds for the two-particle states in two different cases: one or two bosonic modes \textit{with no} Majorana modes; and two Majorana modes with one or two bosonic modes.

Let us first focus on the case where there is only one excitation in the Majorana sector.
The coefficients of the MPS representation of the three-particle Moore-Read state (\ref{coef_MR}) with a single excitation in the Majorana sector, $\chi_{r-1/2}$, are:
\begin{align}
C_{k_{1},k_{2},k_{3}}^{r} & \!\negthinspace=\!\!\frac{1}{\prod_{i}l_{i}!}\!\!\!\sum_{\lambda_{1},\lambda_{2},\lambda_{3}}\!\!\!\!\!\left\langle \!\chi_{r-1/2}\chi_{\lambda_{3}-k_{3}-1/2}\chi_{\lambda_{2}-k_{2}-1/2}\chi_{\lambda_{1}\!-k_{1}\!-1/2}\!\right\rangle \nonumber \\
 & \times\braket{3|V_{-\lambda_{3}-h}V_{-\lambda_{2}-h}V_{-\lambda_{1}-h}|0}\label{coeff_three_part}
\end{align}
where $l_{i}$ is the number of occurrences of $i$ among $k_{1},k_{2},k_{3}$.
If $r=1$ we recover the three-particle densest Moore-Read state coefficients.
Plugging in the results from Appendix \ref{Majo_wick_corr} and \ref{three_V_corr}, we find: 
\begin{align}
C_{k_{1},k_{2},k_{3}}^{r} & =\frac{1}{\prod_{i}l_{i}!}\biggl(\sum_{\lambda_{2}>k_{2}}h^{q}\left(k_{1}-r+1,\lambda_{2},-\lambda_{2}+k_{2}+k_{3}+1\right)\nonumber \\
 & -\sum_{\lambda_{1}>k_{1}}h^{q}\left(\lambda_{1},k_{2}-r+1,-\lambda_{1}+k_{1}+k_{3}+1\right)\nonumber \\
 & +\sum_{\lambda_{1}>k_{1}}h^{q}\left(\lambda_{1},-\lambda_{1}+k_{1}+k_{2}+1,k_{3}-r+1\right)\biggr).\label{coef-3-part-1excit}
\end{align}
Here, $h^{q}$ is a real function and is defined in Eq. \eqref{h^q}.

The above equation (\ref{coef-3-part-1excit}) can be expressed in a more symmetrical form as follows:
\begin{align}
C_{k_{1},k_{2},k_{3}}^{r} & =\frac{\left(-1\right)^{q}}{\prod_{i}l_{i}!}\sum_{\lambda>0}\biggl(h^{q}\left(\lambda+k_{2},k_{1}-r+1,-\lambda+k_{3}+1\right)\nonumber \\
 & -\left(-1\right)^{q}h^{q}\left(\lambda+k_{1},k_{2}-r+1,-\lambda+k_{3}+1\right)\nonumber \\
 & +h^{q}\left(\lambda+k_{1},k_{3}-r+1,-\lambda+k_{2}+1\right)\biggr),\label{mps_coef_3_particles_1majorExcitation}
\end{align}
where we rearrange the summation over $\lambda$ to start at zero for each of the three terms, and permute the arguments of $h^q$ while using the (anti-)symmetry in the arguments, adding a factor of $(-1)^{q}$ for each permutation.
All non-zero coefficients possess angular momentum equal to $k_{1}+k_{2}+k_{3}=3q+r-2$. The densest $N$-particles Moore-Read state has total angular momentum $L=q\left(N-1\right)-\left\lfloor{N/2}\right\rfloor$, where $\left\lfloor \right\rfloor $ denotes the integer part. As a consequence, the Majorana excitation $\chi_{r-1/2}$ raises the overall angular momentum of the three-particle system by $r-1$.

The zero mode condition (\ref{T-ind-assump_3}) and (\ref{Q-ind-assump_3})  can be shown
by directly applying the $T_{J}^{t}$ (and $Q_{J}^{t}$ ) to the three-particle MR state. We focus first on the three-body Hamiltonian part:
\[
T_{J}^{t}\ket{\psi_{3}^{\chi_r}}=T_{J}^{t}\sum_{\left\{ k_{i}\right\} }\frac{\prod_{i}l_{i}!}{3!}C_{k_{1},k_{2},k_{3}}^{r}c_{k_{1}}^{*}c_{k_{2}}^{*}c_{k_{3}}^{*}\ket{0}.
\]
The three annihilation operators in $T_{J}^{t}$ are contracted with
the three creation operators in the $\ket{k_{3},k_{2},k_{1}}$. The
exchange of two of any two arguments of the form factor, $f^{t}$, of
$T_{J}^{t}$, adds a factor-phase $(-1)^{t}$. Therefore we can make the replacement 
$f^{t}(m,n,p)c_{m}c_{n}c_{p}c_{k_{1}}^{*}c_{k_{2}}^{*}c_{k_{3}}^{*}\equiv 3\left(1-(-1)^{q}(-1)^{t}\right)f^{t}(m,n,p)\delta_{p,k_{1}}\delta_{n,k_{2}}\delta_{m,k_{3}}$ in the resulting sum.
As the symmetry of the form factors $(-1)^{t}$ in Eq. \ref{H_3bd} is constrained to be the same as that of the particle statistics $(-1)^{q-1}$, we have $\left(1-(-1)^{q}(-1)^{t}\right)=2$. Consequently, we can express the above equation as follows:
\begin{equation}
\bra{0}T_{J}^{t}\ket{\psi_{3}^{\chi_r}}=\!\!\!\sum_{\substack{m+n+p=J}
}\!\!\!\!\!\!f^{t}(m,n,p)\left(\prod_{i}l_{i}!\right)\,C_{p,n,m}^{r}.\label{Qpsi_3_f_C}
\end{equation}
Observing the commutation properties of $h^{q}$ as detailed in Appendix \ref{three_V_corr} and the symmetry of $f^{t}$, $(-1)^{t}$, we find that we may make the replacement $\left(\prod_{i}l_{i}!\right)f^{t}(m,n,p)C_{p,n,m}^{r}\equiv3f^{t}(m,n,p)\sum_{\lambda=1}^{2q}h^{q}\left(\lambda+n,p-r+1,m-\lambda+1\right)$
inside the sum over $m,n,p$. 
The factor of three arises from the presence of three $h^{q}$ functions in Eq. (\ref{mps_coef_3_particles_1majorExcitation}).
Therefore, we can express the equation above as follows:
\begin{align}
\bra{0}T_{J}^{t}\ket{\psi_{3}^{\chi_r}}&=  3\frac{(-1)^{q}}{\prod_{i}l_{i}!}\sum_{\substack{m+n+p=J}
}\!\!\!\!\!f^{t}(m,n,p)\nonumber \\
 & \!\!\!\!\times\sum_{\lambda=1}^{2q}h^{q}\left(\lambda+n,p-r+1,m-\lambda+1\right)=0. \label{3bd_annih_3part_state}
\end{align}
Appendix \ref{N3_1Majo_mode} demonstrates that the above expression evaluates to zero.

Let's now examine a situation where there exists a single bosonic mode excitation and one Majorana excitation. This analysis will also show that extending the procedure to two or three bosonic modes is straightforward.
The  MPS coefficient for a single excitation in each CFT sector, represented by the superscripts ${r,l}$, can be expressed as:
\begin{align}
C_{k_{1},k_{2},k_{3}}^{r,l} & \!\negthinspace=\!\!\frac{1}{\prod_{i}l_{i}!}\!\!\!\sum_{\lambda_{1},\lambda_{2},\lambda_{3}}\!\!\!\!\!\left\langle \!\chi_{r-1/2}\chi_{\lambda_{3}-k_{3}-1/2}\chi_{\lambda_{2}-k_{2}-1/2}\chi_{\lambda_{1}\!-k_{1}\!-1/2}\!\right\rangle \nonumber \\
 &\qquad\qquad\ \times\braket{3|a_{l}V_{-\lambda_{3}-h}V_{-\lambda_{2}-h}V_{-\lambda_{1}-h}|0}.\label{mps_coef_1maj_1bos}
\end{align}
By utilizing the commutation relation provided in Eq. (\ref{comut_a_n,V_lambda}), we can derive the following expression for the bosonic-sector correlator by commuting $a_l$ to the right:
\begin{align}
\left\langle a_{l}V_{-\lambda_{3}-h}V_{-\lambda_{2}-h}V_{-\lambda_{1}-h}\right\rangle&  =  \sqrt{q}\bigl\langle V_{-\lambda_{3}+l-h}V_{-\lambda_{2}-h}V_{-\lambda_{1}-h}\bigr\rangle + \nonumber \\
\sqrt{q}\bigl\langle V_{-\lambda_{3}-h}V_{-\lambda_{2}+l-h} V_{-\lambda_{1}-h}\bigr\rangle + & \sqrt{q}\bigl\langle V_{-\lambda_{3}-h}V_{-\lambda_{2}-h}V_{-\lambda_{1}+l-h}\bigr\rangle.\label{anVVV}
\end{align}


{ After the substitution of this expression into Eq.(\ref{mps_coef_1maj_1bos}) and shifting summation variables, 
 we can rewrite the $3$-particles MPS coefficient with a single bosonic and a single Majorana excitation as follows:}
\[
C_{k_{1},k_{2},k_{3}}^{r,l}=\sqrt{q}\left(C_{k_{1},k_{2},k_{3}-l}^{r}+C_{k_{1},k_{2}-l,k_{3}}^{r}+C_{k_{1}-l,k_{2},k_{3}}^{r}\right).
\]
Hence, we conclude that the state, with MPS coefficients as defined in Eq. (\ref{mps_coef_1maj_1bos}), can be expressed as follows:
\begin{equation}
\sum_{(k_{i})^3}\!\!C_{k_{1},k_{2},k_{3}}^{r,l}\!\!\ket{k_{3},k_{2},k_{1}}=\sqrt{q}p_{l}\!\!\!\sum_{(k_{i})^3}\!\!C_{k_{1},k_{2},k_{3}}^{r}\!\!\ket{k_{3},k_{2},k_{1}},\label{p_n_on_3part_1majo}
\end{equation}
where $(k_{i})^3$ is defined in \ref{def_k_{i})^N} and 
$p_l$ is the operator $p_{l}=\sum_{k}c_{k+l}^{*}c_{k}$,
whose action  within the variational subspace we thus see to have the same effect as the addition of a bosonic field $a_l$ in the MPS description. It is well known \cite{Mazaheri2015a} that more generally, this operator facilitates the multiplication with the power-sum polynomial $\sum_{i=1}^{N} z^l_i$ in first quantization.

It turns out that the state in Eq. (\ref{p_n_on_3part_1majo}) is annihilated by $T_{J}^{t}$. This happens because 
\begin{equation}
    [T_{J}^{t},p_{l}] = \sum_{m+n+p=J-l}g^{t,l}\left(m,n,p\right)c_{m}c_{n}c_{p} \label{[T,p]}
\end{equation}
Here, the form factor $g^{t,l}(m,n,p)$, defined as $g^{t,l}(m,n,p)=f^{t}(m,n,p+l)+f^{t}(m,n+l,p)+f^{t}(m+l,n,p)$, has the same symmetry as $f^t(m,n,p)$, and likewise depends only on the variables $\left(m-n\right),\ \left(m-p\right),$ and $ \left(n-p\right)$. The operator on the left-hand side of equation \ref{[T,p]} thus closely resembles $T_{J}^{t}$, differing only in the form factor and the decreased total angular momentum from $J$ to $J-l$.
In equation (\ref{3bd_annih_3part_state}), we had previously proven that $T_{J}^{t}$ annihilates three-particle states with one Majorana mode, for an arbitrary value of $J$. Since $g^{t,l}$ shares the same essential properties with $f^t$, the same demonstration can be performed by substituting $f^t$ with $g^{t,l}$.
Therefore, when $T_{J}^{t}$ 
is applied to the state in question, the result is zero.
This argument can be repeated systematically in the presence of multiple bosonic modes excitations.

We can apply the same methodology used for the case with one Majorana mode to obtain the outcome for the state of three particles with three excitations in the Majorana sector, specifically, $\chi_{r_{3}-1/2},\ \chi_{r_{2}-1/2},\ \text{and}\ \chi_{r_{1}-1/2}$.
This state's MPS coefficient is
\begin{align}
C_{k_{1},k_{2},k_{3}}^{r_{1},r_{2},r_{3}}=\frac{1}{\prod_{i}l_{i}!}\bigg( & h^{q}\left(k_{1}-r_{1}+1,k_{2}-r_{2}+1,k_{3}-r_{3}+1\right)\nonumber \\
 & -(-1)^{q}h^{q}\left(k_{2}-r_{1}+1,k_{1}-r_{2}+1,k_{3}-r_{3}+1\right)\nonumber \\
 & +\left(k_{1},k_{2},k_{3}\right)\rightarrow\left(k_{2},k_{3},k_{1}\right)\nonumber \\
 & +\left(k_{1},k_{2},k_{3}\right)\rightarrow\left(k_{3},k_{1},k_{2}\right)\bigg). \label{MPS-coef-3part-3majo}
\end{align}

The operators $T_{J}^{t}$ can be applied to the MR state with three excitations in the Majorana sector, similar to what was done in equation (\ref{Qpsi_3_f_C}). However, this time, the resulting replacement inside sums over of $m,n,$ and $p$ is $\prod_{i}l_{i}!f^{t}\!(m,n,p)C_{p,n,m}^{r_{1},r_{2},r_{3}}\equiv6f^{t}\!(m,n,p)h^{q}\left(\!p-r_{1}+\!1,\!n-r_{2}+\!1,\!m-r_{3}+\!1\right)$. This yields:
\begin{align}
\bra{0}T_{J}^{t}\ket{\psi_{3}^{\chi_{r_{1}},\chi_{r_{2}},\chi_{r_{3}}}}\! &=  6\!\!\!\!\!\!\sum_{\substack{m+n+p=J}
}\!\!\!\!\!\!f^{t}(m,n,p)\nonumber \\
 &\!\!\!\!\!\!\!\!\!\!\!\!\!\times h^{q}\left(p-r_{1}+1,n-r_{2}+1,m-r_{3}+1\right)=0
\end{align}
Appendix \ref{N3_3Majo_mode} demonstrates that the above equation evaluates to zero.

It remains again to consider the introduction of bosonic modes $a_n$.
Introducing any number of $a_{n}$ modes to the bosonic sector will solely result in negative integers being added to each index of the vertex operators. For instance, the first term in equation (\ref{MPS-coef-3part-3majo}) would be a linear combination of terms $h^{q}\left(k_{1}-r_{1}-n_{1}+1,k_{2}-r_{2}-n_{2}+1,k_{3}-r_{3}-n_{3}+1\right)$.
It is evident that we can redefine $r_{i}+n_{i}\rightarrow r_{i}$. Given that  the $r_{i}'s$ are essentially arbitrary, the MPS coefficients that are written in Eq. (\ref{MPS-coef-3part-3majo}) above already account for the addition of any number of edge excitations in the bosonic sector when considering three excitations in the Majorana sector.

\subsection{The induction beginning: two-particle zero mode property}

It remains to fully demonstrate the application of the Theorem to the 2-body Hamiltonian $H_{\frac{1}{q}}^{\text{2bd}}$. To this end, we must also establish ${\cal W}_k \subset H^0_k$ for this case, where, in this sub-section, the $H^0_N$ denote the
zero mode spaces of $H_{\frac{1}{q}}^{\text{2bd}}$.
 To do so, we start with the general MR MPS state of two particles, which can be expressed as
\begin{equation}
\ket{\psi_{2}^{a_{n}...,\chi_{l}...}}=\sum_{k_{1},k_{2}}C_{k_{1},k_{2}}^{\{r,n\}}\ket{k_{1},k_{2}},\label{MR_MPS_2particles}
\end{equation}
where the indices ${r,n}$ represent the edge excitations in the Majorana and bosonic sector, respectively. It is important to note that any MR state of two particles can be written as a linear combination of states with two or no Majorana excitations up to
two arbitrary bosonic excitations.
We first demonstrate the zero mode condition for a state with two excitations in the Majorana sector. The corresponding MPS-coefficient is:
\begin{align}
C_{k_{1},k_{2}}^{r_{1},r_{2}}=\!\! & \sum_{\lambda_{1},\lambda_{2}}\left\langle \!\chi_{r_{1}-1/2}\chi_{r_{2}-1/2}\chi_{\lambda_{2}-k_{2}-1/2}\chi_{\lambda_{1}-k_{1}\!-1/2}\!\right\rangle \nonumber \\
 & \qquad\times\braket{2|V_{-\lambda_{2}-h}V_{-\lambda_{1}-h}|0}.
\end{align}
Utilizing equation (\ref{majorana_2modes_2excitations}) in Appendix \ref{Majo_wick_corr} for the correlator of Majorana modes, we obtain:
\begin{align}
C_{k_{1},k_{2}}^{r_{1},r_{2}}= & \braket{2|V_{-\left(k_{2}-r_{2}+1\right)-h}V_{-\left(k_{1}-r_{1}+1\right)-h}|0}\nonumber \\
 & \ -\braket{2|V_{-\left(k_{2}-r_{1}+1\right)-h}V_{-\left(k_{1}-r_{2}+1\right)-h}|0}.
\end{align}
Note that each of these MPS coefficients (where $r_i \geq 1$) are exactly in the same form as the Laughlin $2$-particles MPS coefficients described in Ref. \onlinecite{Schossler2022}, where the zero mode condition has already been proven for the same two-body Hamiltonian. Therefore,
\begin{equation}
Q_{J}^{m}\ket{\psi_{2}^{\chi{r_{1}},\chi{r_{2}}}}=0.\label{Q-2part-2majo}
\end{equation}
By analogy with the MR state with three particles, we can also redefine $r_{1}$ and $r_{2}$ upon introduction of one or two bosonic excitations. Using the result from equation \ref{Q-2part-2majo} shows that the $2$-particle MR state with two Majorana and two bosonic excitations then satisfies the zero mode condition for the two-body Hamiltonian $Q_{J}^{m}$.

In the following, we will prove that this statement remains valid for the MR state of two particles and exclusively bosonic excitations.
The MPS of this state is
\begin{align}
C_{k_{1},k_{2}}^{n,l}\!\! & =\!\!\!\!\!\sum_{\lambda_{1},\lambda_{2}}\left\langle \chi_{\lambda_{2}-k_{2}-1/2}\chi_{\lambda_{1}-k_{1}\!-1/2}\!\right\rangle \braket{2|V_{-\lambda_{2}+n-h}V_{-\lambda_{1}+l-h}|0}\nonumber \\
 & =\!\!\sum_{\lambda=0}^{k_{1}}\braket{2|V_{-\left(-\lambda+k_{1}+k_{2}+1\right)+n-h}V_{-\lambda+l-h}|0}.
\end{align}
This equation's second line is obtained through the contraction of two Majorana fields: $\left\langle \chi_{\lambda_{2}-k_{2}-1/2}\chi_{\lambda_{1}-k_{1}\!-1/2}\!\right\rangle =\delta_{\lambda_{1}-k_{1}+\lambda_{2}-k_{2},1}$, under the additional  constrain that $\lambda_{2}-k_{2}\geq1$.
To calculate the correlator of two vertex modes, we can use the last equation from Appendix B of Ref. \onlinecite{Schossler2022}: $\braket{2,0|V_{-a-h}V_{-b-h}|0}=\delta_{q,a+b}\left(-1\right)^{b}{q \choose b}$. Consequently, we obtain:
\begin{equation}
    C_{k_{1},k_{2}}^{n,l}\!\!=\delta_{q-1+n+l,k_{1}+k_{2}}\left(-1\right)^{k_{1}-l}{q-1 \choose k_{1}-l}.\label{coef-2part-2boson}
\end{equation}
To obtain this closed form, we utilized the identity of the alternating sums and differences of binomial coefficients up to $k$, which states that
\begin{equation}
\sum_{\lambda=0}^{k}\left(-1\right)^{\lambda}{q \choose \lambda}=\left(-1\right)^{k}{q-1 \choose k}.\label{alternating_sum_identity}
\end{equation}

Plugging in the coefficient from equation (\ref{coef-2part-2boson}) into equation (\ref{MR_MPS_2particles}), we obtain
\begin{align}
 & \bra{0}Q_{J}^{m}\ket{\psi_{2}^{a_n,a_l}}\!=\!\!\!\!\!\sum_{x=-J/2}^{J/2}x^{m}\frac{C_{k_{1},k_{2}}^{n,l}\!\!}{2}\bra{0}c_{J/2-x}c_{J/2+x}c_{k_{1}}^{*}c_{k_{2}}^{*}\ket{0}\\
 & =\!\!\!\sum_{x=-J/2}^{J/2}\!\!\!x^{m}\frac{C_{k_{1},k_{2}}}{2}\Big(\delta_{J/2+x,k_{1}}\delta_{J/2-x,k_{2}}\nonumber \\
 &\qquad\qquad\qquad\qquad+\left(-1\right)^{q-1}\delta_{J/2-x,k_{1}}\delta_{J/2+x,k_{2}}\Big)\\
 & =\!\!\!\!\!\sum_{x=-J/2}^{J/2}\!\!\!\!\!x^{m}\delta_{q-1+n+l,k_{1}+k_{2}}\left(-1\right)^{k_{1}-l}{q-1 \choose k_{1}-l}\delta_{J/2+x,k_{1}}\delta_{J/2-x,k_{2}}\\
 & =\delta_{q-1+n+l,J}\!\!\!\!\!\sum_{x=-l}^{q-1+n}\left(x-\frac{q-1+n-l}{2}\right)^{m}\left(-1\right)^{x}{q-1 \choose x}\\
 & =0\ \text{if}\ m<q-1
\end{align}
Here, in going to the final line we used Eq. (\ref{relation_Ruiz1996}), cf. Ref. \onlinecite{Ruiz1996}.
 The $Q_J^m$ operators in the Moore-Read two-body Hamiltonian (equation \ref{H_2bd}) are limited to $m<q-1$. This implies that the $2$-particle Moore-Read state with two or fewer ($n,\ l$ can assume zero value) bosonic excitations satisfies the zero mode condition. Hence, we have shown that the two-body Hamiltonian in equation \ref{H_2bd} stabilizes the Moore-Read state with any number of excitations in both CFT sectors.

Taken together, we have now shown using the Theorem that the MPS states lie in the intersection of the zero mode spaces of both $H_{\frac{1}{q}}^{\text{2bd}}$ and $H_{\frac{1}{q}}^{\text{3bd}}$, thus, the MPS variational spaces ${\cal W}_N$ comprise zero modes of the full Hamiltonian $H_{\frac{1}{q}}$, \Eq{Hfull}.

\section{Conclusion\label{conclusion}}


We have proven a Theorem that may establish the existence of a frustration free parent Hamiltonian in situations where where the direct application of such a Hamiltonian to trial ground state wave functions presents significant challenges. A major motivation for this Theorem derives from scenarios in which trial wave functions take the form of infinite bond dimension MPS generated by conformal field theory, as are well-known to emerge in the realm of fractional quantum Hall physics.
The neeed for such a Theorem is evident from the fact that traditional MPS with finite bond dimensions have historically been well-suited for constructing parent Hamiltonians. Conversely, the connection between these Hamiltonians and infinite-dimensional MPS has remained somewhat unclear in the fractional quantum Hall literature.
We have extensively discussed the utility of our theorem in the context of the non-Abelian Moore-Read state. We submit that this utility extends beyond lowest-Landau level conformal field theory inspired wave functions. For instance, in mixed Landau levels, there exists an abundance of promising parton states\cite{2bar11,wu2017new,Parton_antiPfaffian,1bar2bar111,fwave,PhysRevB.103.085303,PhysRevResearch.3.033087,PhysRevB.103.155103,Balram:2021opn,PhysRevB.105.L241403} that should meet the same conditions required by our Theorem whenever a parent Hamiltonian exists. There is, moreover, no particular reason why our Theorem should be limited in application to the fractional quantum Hall regime or, more generally, to quasi-one-dimensional Hamiltonians. We are thus hopeful that the progress made in this paper may open up new avenues for the creation of solvable models.

\begin{acknowledgments}
Work by A.S. and M.S. has been supported by the National Science Foundation under Grant No. DMR-2029401.
L.C. is supported by NSFC Grant No. 12004105. 
A.S. is grateful for insightful discussions {with} J.I. Cirac. 
\end{acknowledgments}

\onecolumngrid
\appendix


{\section{{The product of} $T_{r}$ and $\hat{N}$}\label{ident_show}
In this appendix, we demonstrate Eq. \ref{genTident}. This identity can be simply obtained by calculating the {product} between the operators $T_{r}$ and $\hat{N}$ as follows:
\begin{align}
T_{r}\hat{N} & =\sum_{j_{1}\dots j_{k}}\sum_{j}\eta_{j_{1}\dots j_{k}}^{r}c_{j_{1}}\dots c_{j_{k}}c_{j}^{\ast}c_{j}\nonumber\\
  =&\sum_{j_{1}\dots j_{k}}\sum_{j}\eta_{j_{1}\dots j_{k}}^{r}c_{j_{1}}\dots\left[\delta_{j_{k},j}+(-1)^{\xi}c_{j}^{\ast}c_{j_{k}}\right]c_{j}\nonumber\\
 =&\sum_{j_{1}\dots j_{k}}\eta_{j_{1}\dots j_{k}}^{r}c_{j_{1}}\dots c_{j_{k}} +(-1)^{\xi}\sum_{j_{1}\dots j_{k}}\sum_{j}\eta_{j_{1}\dots j_{k}}^{r}c_{j_{1}}\dots c_{j}^{\ast}c_{j_{k}}c_{j}\nonumber\\
 & \vdots \nonumber\\
 =&k\sum_{j_{1}\dots j_{k}}\eta_{j_{1}\dots j_{k}}^{r}c_{j_{1}}\dots c_{j_{k}}+(-1)^{k\xi}\sum_{j_{1}\dots j_{k}}\sum_{j}\eta_{j_{1}\dots j_{k}}^{r}c_{j}^{\ast}c_{j_{1}}\dots c_{j_{k}}c_{j}\nonumber\\
 =&kT_{r}+(-1)^{k\xi}\sum_{j}c_{j}^{\ast}T_{r}c_{j}.
\end{align}
In the second line, we used the canonical commutation relation between $c_{j}^{\ast}$ and $c_{j_{k}}$. In the fourth line, we repeatedly applied the commutation relation for $c_{j}^{\ast}$ and $c_{j_{i}}$, where $i=1,\dots,k-1$. It's important to note that $c_{j_{i}}c_{j} = (-1)^{\xi}c_{j}c_{j_{i}}$, ensuring that each term with the string of operators $c_{j_{1}}\dots c_{j_{k}}$ has the phase $(-1)^{\xi}$ canceled out. The final result is derived by invoking the definition of $T_r$.
}

\section{Wick expansion of correlators}\label{Majo_wick_corr}

This appendix presents the Wick expansions of the correlators of the modes in the Majorana sector that are used in the main text.
These are, in particular, related to MPS coefficients for two and three particles with considering one, two, or three Majorana excitations, respectively. Furthermore, we demonstrate that a Moore-Read MPS
with $N$ particles and
with any number of excitations 
 in the Majorana sector  can be expressed as a linear combination of such MPS with only $N$ excitations.

The correlator of the three particles Majorana modes with one excitation is given by:
\begin{align}
\braket{0|\chi_{r_{1}-1/2}\chi_{l_{3}-1/2}\chi_{l_{2}-1/2}\chi_{l_{1}-1/2}|0} =& \braket{0|\chi_{r_{1}-1/2}\chi_{l_{3}-1/2}|0}\braket{0|\chi_{l_{2}-1/2}\chi_{l_{1}-1/2}|0}\nonumber \\
 & -\braket{0|\chi_{r_{1}-1/2}\chi_{l_{2}-1/2}|0}\braket{0|\chi_{l_{3}-1/2}\chi_{l_{1}-1/2}|0}\nonumber \\
 & +\braket{0|\chi_{r_{1}-1/2}\chi_{l_{1}-1/2}|0}\braket{0|\chi_{l_{3}-1/2}\chi_{l_{2}-1/2}|0}\\
 = &\, \theta(l_{2}-1)\delta_{l_{3},-r_{1}+1}\delta_{l_{1},-l_{2}+1}-\theta(l_{1}-1)\delta_{l_{2},-r_{1}+1}\delta_{l_{1},-l_{3}+1}\nonumber\\
 & +\theta(l_{1}-1)\delta_{l_{1},-r_{1}+1}\delta_{l_{2},-l_{3}+1}
\end{align}
where $r_1\geq 1$, and
\begin{equation}
    \theta(x)=
\begin{cases}
0, & \text{if } x < 0, \\
1, & \text{if } x \geq 0.
\end{cases}
\end{equation}
is the Heaviside function.
{This effectively restricts the Majorana mode indexes to:
\begin{align}
l_{2} & \geq1,\ \text{for the first term}\nonumber \\
l_{1} & \geq1,\ \text{for the second term}\nonumber \\
l_{1} & \geq1,\ \text{for the third term.}\label{restrictions_majorana}
\end{align}}

The correlator of the three particles Majorana modes with three excitations is:
\begin{align}
\braket{0|\chi_{r_{3}-1/2}\chi_{r_{2}-1/2}\chi_{r_{1}-1/2}\chi_{l_{1}-1/2}\chi_{l_{2}-1/2}\chi_{l_{3}-1/2}|0} & =\braket{0|\chi_{r_{1}-1/2}\chi_{l_{1}-1/2}|0}\braket{0|\chi_{r_{2}-1/2}\chi_{l_{2}-1/2}|0}\braket{0|\chi_{r_{3}-1/2}\chi_{l_{3}-1/2}|0}\nonumber \\
 & -\braket{0|\chi_{r_{1}-1/2}\chi_{l_{2}-1/2}|0}\braket{0|\chi_{r_{2}-1/2}\chi_{l_{1}-1/2}|0}\braket{0|\chi_{r_{3}-1/2}\chi_{l_{3}-1/2}|0}\nonumber \\
 & +\braket{0|\chi_{r_{1}-1/2}\chi_{l_{2}-1/2}|0}\braket{0|\chi_{r_{2}-1/2}\chi_{l_{3}-1/2}|0}\braket{0|\chi_{r_{3}-1/2}\chi_{l_{1}-1/2}|0}\nonumber \\
 & -\braket{0|\chi_{r_{1}-1/2}\chi_{l_{3}-1/2}|0}\braket{0|\chi_{r_{2}-1/2}\chi_{l_{2}-1/2}|0}\braket{0|\chi_{r_{3}-1/2}\chi_{l_{1}-1/2}|0}\nonumber \\
 & +\braket{0|\chi_{r_{1}-1/2}\chi_{l_{3}-1/2}|0}\braket{0|\chi_{r_{2}-1/2}\chi_{l_{1}-1/2}|0}\braket{0|\chi_{r_{3}-1/2}\chi_{l_{2}-1/2}|0}\nonumber \\
 & -\braket{0|\chi_{r_{1}-1/2}\chi_{l_{1}-1/2}|0}\braket{0|\chi_{r_{2}-1/2}\chi_{l_{3}-1/2}|0}\braket{0|\chi_{r_{3}-1/2}\chi_{l_{2}-1/2}|0}\\
 & =\delta_{-r_{1},l_{1}-1}\delta_{-r_{2},l_{2}-1}\delta_{-r_{3},l_{3}-1}-\delta_{-r_{1},l_{2}-1}\delta_{-r_{2},l_{1}-1}\delta_{-r_{3},l_{3}-1}\nonumber \\
 & \ \ +\delta_{-r_{1},l_{2}-1}\delta_{-r_{2},l_{3}-1}\delta_{-r_{3},l_{1}-1}-\delta_{-r_{1},l_{3}-1}\delta_{-r_{2},l_{2}-1}\delta_{-r_{3},l_{1}-1}\nonumber \\
 & \ \ +\delta_{-r_{1},l_{3}-1}\delta_{-r_{2},l_{1}-1}\delta_{-r_{3},l_{2}-1}-\delta_{-r_{1},l_{1}-1}\delta_{-r_{2},l_{3}-1}\delta_{-r_{3},l_{2}-1}
\end{align}
where $r_{i}\geq1$, in the terms above.

The correlator of the two particles Majorana modes with two excitations is:
\begin{align}
\left\langle \!\chi_{r_{1}-1/2}\chi_{r_{2}-1/2}\chi_{l_{2}-1/2}\chi_{l_{1}\!-1/2}\!\right\rangle  & =\braket{0|\chi_{r_{1}-1/2}\chi_{l_{1}-1/2}|0}\braket{0|\chi_{r_{2}-1/2}\chi_{l_{2}-1/2}|0}\nonumber \\
 & -\braket{0|\chi_{r_{1}-1/2}\chi_{l_{2}-1/2}|0}\braket{0|\chi_{r_{2}-1/2}\chi_{l_{1}-1/2}|0}\\
 & =\delta_{l_{1},-r_{1}+1}\delta_{l_{2},-r_{2}+1}-\delta_{l_{2},-r_{1}+1}\delta_{l_{1},-r_{2}+1}\label{majorana_2modes_2excitations}
\end{align}
As mentioned before, the equations presented above require that $r_i \geq 1$ since $\bra{0}\chi_{r_{i}-1/2}$ would otherwise be zero.

We will now demonstrate that any number of excitations in the three-particle Moore-Read state can be expressed in terms of one or three excitations. 
Suppose we add two more excitations, $\chi_{r_{4}-1/2}$ and $\chi_{r_{5}-1/2}$, to the correlator of the three particles Majorana modes with three excitations. Because there are three modes from three particles and five excitation modes in total, when we apply Wick's theorem, each term in the expansion will contain a factor that involves the correlator of two Majorana modes with excitation indices, $\left\langle \chi_{r_{i}-1/2},\chi_{r_{j}-1/2}\right\rangle$.
However, because of the anticommutation relation for the $\chi_{r}$'s and the relations (\ref{conditions_modes_majo}), this factor will be either zero or one.
Thus, we have demonstrated that any combination of excitations can always be expressed as a linear combination of three excitations or fewer. Moreover, for $N$ particles, $N$ Majorana mode excitations or fewer are required.

\section{Three-vertex correlator}\label{three_V_corr}

In this appendix, we will compute the correlator of three vertex operators and express it in terms of a simple
real function, $h^{q}$. This result is useful in demonstrating the zero mode property of the Moore-Read state with three particles. The correlator of three vertex modes is:
\begin{align}
\braket{3|V_{-a-h}V_{-b-h}V_{-c-h}|0} & =\frac{1}{\left(2\pi i\right)^{3}}\ointop\frac{dz_{1}}{z_{1}^{a+1}}\ointop\frac{dz_{2}}{z_{2}^{b+1}}\ointop\frac{dz_{3}}{z_{3}^{c+1}}\braket{3|V\left(z_{1}\right)V\left(z_{2}\right)V\left(z_{3}\right)|0}\\
 & =\frac{1}{\left(2\pi i\right)^{3}}\ointop\frac{dz_{1}}{z_{1}^{a+1}}\ointop\frac{dz_{2}}{z_{2}^{b+1}}\ointop\frac{dz_{3}}{z_{3}^{c+1}}\left(z_{1}-z_{2}\right)^{q}\left(z_{1}-z_{3}\right)^{q}\left(z_{2}-z_{3}\right)^{q}\\
 & =\sum_{k_{1},k_{2},k_{3}=0}^{q}\left(-1\right)^{k_{1}+k_{2}+k_{3}}{q \choose k_{1}}{q \choose k_{2}}{q \choose k_{3}}\ointop\frac{dz_{1}}{2\pi i}\ointop\frac{dz_{2}}{2\pi i}\ointop\frac{dz_{3}}{2\pi i}\frac{z_{1}^{2q-k_{1}-k_{2}}}{z_{1}^{a+1}}\frac{z_{2}^{q+k_{1}-k_{3}}}{z_{2}^{b+1}}\frac{z_{3}^{k_{2}+k_{3}}}{z_{3}^{c+1}}\\
 & =\sum_{k_{1},k_{2},k_{3}=0}^{q}\left(-1\right)^{k_{1}+k_{2}+k_{3}}{q \choose k_{1}}{q \choose k_{2}}{q \choose k_{3}}\delta_{2q-k_{1}-k_{2},a}\delta_{q+k_{1}-k_{3},b}\delta_{k_{2}+k_{3},c}\\
 & =\delta_{3q,a+b+c}\sum_{k=0}^{q}\left(-1\right)^{k+c}{q \choose k}{q \choose a+k-q}{q \choose b-k}\label{3_particle_vertex_mode}
\end{align}
The binomial expansion was applied in the second line to obtain the expression in the third line. It is important to note that the following restrictions must be imposed to ensure non-zero results:
\begin{equation}
0\leq a,b,c\leq2q.
\end{equation}
These restrictions arise due to the addition of a background charge of three units in the out-state and zero units in the in-state, which constrains the angular momentum associated with any vertex mode above \cite{Schossler2022}.
To reduce notation, we define:
\begin{equation}
h^{q}\left(a,b,c\right)\doteq\braket{3|V_{-a-h}V_{-b-h}V_{-c-h}|0}.\label{h^q}
\end{equation}
The commutation properties of the modes of the vertex operator are also reflected by the (anti-)symmetry of the real function $h^{q}$ as  follows: $h^{q}\left(a,b,c\right)=(-1)^{q}h^{q}\left(b,a,c\right)=(-1)^{q}h^{q}\left(c,b,a\right)=h^{q}\left(c,a,b\right)$.

\section{$N=3$ zero mode property with single Majorana mode}\label{N3_1Majo_mode}
In this appendix, we demonstrate that the three-body operator $T_{J}^{t}$ annihilates a Moore-Read state with three particles and a single Majorana mode excitation. In this proof, we compute the overlap of the resulting zero-particle state $T_{J}^{t}\ket{\psi_{3}^{r}}$ with the vacuum. This overlap is:
\begin{align}
\bra{0}\!T_{J}^{t}\!\ket{\psi_{3}^{\chi_r}}\!\!= & 3\delta_{3q+r-2,J}\!\!\!\!\!\sum_{\substack{m+n+p=J}
}\!\sum_{\lambda=1}^{2q}\left(m-n\right)^{t_1}\left(m-p\right)^{t_2}\left(n-p\right)^{t_3}\sum_{k=0}^{q}\binom{q}{k}(-1)^{k-\lambda+m+1}\binom{q}{k+n-q+\lambda}\binom{q}{-k+p-r+1}\\
= & 3\!\!\!\!\!\!\!\!\sum_{0\leq k,p\leq q}\!\!\!\!\sum_{n=0}^{3q+r-2}\!\!\!\!(-1)^{k-p}(3q-\!k-\!2n-\!p-\!1)^{t_1}\!(n-\!k-\!p-\!r+\!1)^{t_2}\!(3q-\!2k-\!n-\!2p-\!r)^{t_3}\!\binom{q}{k}\!\binom{q}{p}\!\!\!\sum_{\lambda=1+n+k-q}^{q}\!\!\!\!\!\!(-1)^{-\lambda}\binom{q}{\lambda}\\
= & -3\!\!\!\!\!\!\!\!\sum_{0\leq k,p\leq q}\!\!\!\!\sum_{n=0}^{3q+r-2}\!\!\!\!(-1)^{q-p-n}(3q-k-2n-p-1)^{t_1}\!(n-k-p-r+1)^{t_2}\!(3q-2k-n-2p-r)^{t_3}\!\binom{q}{k}\!\binom{q}{p}\!\binom{q-1}{k\!+n-\!q}\\
= & -3\!\!\!\!\!\!\!\!\sum_{0\leq k,p\leq q}\!\sum_{n=0}^{q-1}(-1)^{k+n+p}(k-2n-p+q-1)^{t_1}\!(q+n-2k-p-r+1)^{t_2}\!(2q-k-n-2p-r)^{t_3}\!\binom{q-1}{n}\!\binom{q}{p}\!\binom{q}{k}\label{Qpsi_3_4}\\
= & 0\ \ \diamondsuit
\end{align}
We performed a change of variables on $p\rightarrow p + r - 1 + k$ and $\lambda \rightarrow \lambda -1 - n - k + q$ in the second line 
- the range of the sums are changed using that the binomial coefficient is 0 if its lower index is negative, and following to restrict the range of the sum.
In the third line, we used the identity 
\begin{equation}
    \sum_{\lambda=1+j}^{q}\!(-1)^{-\lambda}\binom{q}{\lambda}=-(-1)^{-j}\binom{q-1}{j}.
\end{equation}
One can derive this identity by splitting the Kronecker delta sum realization, $\delta_{q,0}=\sum_{\lambda=0}^{q}(-1)^{-\lambda}\binom{q}{\lambda}$, into two parts: from $0$ to $j$ and from $j+1$ to $q$. Then, by applying the alternating sum and difference of binomial coefficient up to $k$ identity (Eq. (\ref{alternating_sum_identity})), the desired result is obtained.
In the fourth line, we substituted the variable $n$ and adjusted the limits of the sum.
The final result, in the fifth line, was obtained using the relation\cite{Ruiz1996}
\begin{equation}
\sum_{i=0}^{\beta}(-1)^{i}{\beta \choose i}i^{\alpha}=0\quad\text{for }0\leqslant\alpha<\beta ,\label{relation_Ruiz1996}
\end{equation}
Expanding the polynomial in equation (\ref{Qpsi_3_4}) into monomials on the variables $k$, $n$, and $p$, we observe that for each monomial, at least one of the powers of $k$, $n$, or $p$ is less than $q$ or $q-1$, since the maximum power of the polynomial is $t=3q-3$.
Using the relation in Eq. (\ref{relation_Ruiz1996}), we can conclude that at least one sum in $k,\ n,\text{ or }p$ is zero. Therefore, the overlap above is zero.

\section{$N=3$ zero mode property with three Majorana modes}\label{N3_3Majo_mode}
In this appendix, we demonstrate that the three-body operator $T_{J}^{t}$ annihilates a Moore-Read state with three particles and three Majorana modes excitations. In this proof, we compute the overlap of the resulting zero-particle state $T_{J}^{t}\ket{\psi_{3}^{r}}$ with the vacuum. This overlap is:
\begin{align}
\!\!\!\bra{0}\!T_{J}^{t}\!\ket{\psi_{3}^{\chi_{r_{1}},\chi_{r_{2}},\chi_{r_{3}}}}\!&=  6\delta_{J,3q+r_{1}+r_{2}+r_{3}-3}\!\!\sum_{k=0}^{q}\sum_{\substack{m+n+p=J}
}\!\!\!\!\!\!\left(m\!- n\right)^{t_1}\left(m\!- p\right)^{t_2}\left(n- p\right)^{t_3}\left(-1\right)^{k+ m- r_{3}+1}{q \choose k}{q \choose p- r_{1}+1+ k- q}\!{q \choose n- r_{2}+1- k}\\
&\!\!\!= 6\delta_{J,3q+r_{1}+r_{2}+r_{3}-3}\sum_{k=0}^{q}\sum_{p=-r_{1}+1-k-q}^{J-r_{1}+1+k-q}\sum_{n=-r_{2}+1-k}^{J-r_{2}+1-k}\sum_{m=0}^{J}\binom{q}{k}\binom{q}{n}\binom{q}{p}(-1)^{k+m-r_{3}+1}(-k+m-n-r_{2}+1)^{t_1}\nonumber \\
 & \qquad\qquad\qquad\qquad\qquad\qquad\qquad\qquad\qquad\qquad\times(k+m-p-q-r_{1}+1)^{t_2}(2k+n-p-q-r_{1}+r_{2})^{t_3}\\
&\!\!\!= 6\!\!\!\!\!\!\!\!\sum_{0\leq k,p,n\leq q}\!\binom{q}{k}\!\binom{q}{n}\!\binom{q}{p}\!(-1)^{k+\!n+\!p}(2k+n-p-q-r_{1}+r_{2})^{t_1}\!(k-\!n-2p+q-r_{1}+r_{3})^{t_2}\!(2q-k-2n-p-r_{2}+r_{3})^{t_3}\label{Tpsi3-3}\\
&\!\!\!= 0\ \ \diamondsuit
\end{align}
We performed a change of variables in the first line and then used the binomial factor to restrict the range of the sum to obtain the second line. The result in the third line was obtained using a similar argument as in Appendix \ref{N3_1Majo_mode}.
Expanding the polynomial in equation (\ref{Tpsi3-3}) into monomials on the variables $k$, $n$, and $p$, we observe that for each monomial, at least one of the powers of $k$, $n$, or $p$ is less than $q$ or $q-1$, since the maximum power of the polynomial is $t=3q-3$.
Using the relation in Eq. (\ref{relation_Ruiz1996}), we can conclude that at least one sum in $k,\ n,\text{ or }p$ is zero. Therefore, the overlap above is zero.


\twocolumngrid

 \bibliographystyle{apsrev4-1}
\addcontentsline{toc}{section}{\refname}\bibliography{lib}

\begin{thebibliography}{84}%
\makeatletter
\providecommand \@ifxundefined [1]{%
 \@ifx{#1\undefined}
}%
\providecommand \@ifnum [1]{%
 \ifnum #1\expandafter \@firstoftwo
 \else \expandafter \@secondoftwo
 \fi
}%
\providecommand \@ifx [1]{%
 \ifx #1\expandafter \@firstoftwo
 \else \expandafter \@secondoftwo
 \fi
}%
\providecommand \natexlab [1]{#1}%
\providecommand \enquote  [1]{``#1''}%
\providecommand \bibnamefont  [1]{#1}%
\providecommand \bibfnamefont [1]{#1}%
\providecommand \citenamefont [1]{#1}%
\providecommand \href@noop [0]{\@secondoftwo}%
\providecommand \href [0]{\begingroup \@sanitize@url \@href}%
\providecommand \@href[1]{\@@startlink{#1}\@@href}%
\providecommand \@@href[1]{\endgroup#1\@@endlink}%
\providecommand \@sanitize@url [0]{\catcode `\\12\catcode `\$12\catcode
  `\&12\catcode `\#12\catcode `\^12\catcode `\_12\catcode `\%12\relax}%
\providecommand \@@startlink[1]{}%
\providecommand \@@endlink[0]{}%
\providecommand \url  [0]{\begingroup\@sanitize@url \@url }%
\providecommand \@url [1]{\endgroup\@href {#1}{\urlprefix }}%
\providecommand \urlprefix  [0]{URL }%
\providecommand \Eprint [0]{\href }%
\providecommand \doibase [0]{http://dx.doi.org/}%
\providecommand \selectlanguage [0]{\@gobble}%
\providecommand \bibinfo  [0]{\@secondoftwo}%
\providecommand \bibfield  [0]{\@secondoftwo}%
\providecommand \translation [1]{[#1]}%
\providecommand \BibitemOpen [0]{}%
\providecommand \bibitemStop [0]{}%
\providecommand \bibitemNoStop [0]{.\EOS\space}%
\providecommand \EOS [0]{\spacefactor3000\relax}%
\providecommand \BibitemShut  [1]{\csname bibitem#1\endcsname}%
\let\auto@bib@innerbib\@empty
\bibitem [{\citenamefont {Moore}\ and\ \citenamefont {Read}(1991)}]{Moore1991}%
  \BibitemOpen
  \bibfield  {author} {\bibinfo {author} {\bibfnamefont {G.}~\bibnamefont
  {Moore}}\ and\ \bibinfo {author} {\bibfnamefont {N.}~\bibnamefont {Read}},\
  }\href {\doibase 10.1016/0550-3213(91)90407-O} {\bibfield  {journal}
  {\bibinfo  {journal} {Nuclear Physics B}\ }\textbf {\bibinfo {volume}
  {360}},\ \bibinfo {pages} {362} (\bibinfo {year} {1991})}\BibitemShut
  {NoStop}%
\bibitem [{\citenamefont {Bandyopadhyay}\ \emph {et~al.}(2020)\citenamefont
  {Bandyopadhyay}, \citenamefont {Ortiz}, \citenamefont {Nussinov},\ and\
  \citenamefont {Seidel}}]{Bandyopadhyay2020}%
  \BibitemOpen
  \bibfield  {author} {\bibinfo {author} {\bibfnamefont {S.}~\bibnamefont
  {Bandyopadhyay}}, \bibinfo {author} {\bibfnamefont {G.}~\bibnamefont
  {Ortiz}}, \bibinfo {author} {\bibfnamefont {Z.}~\bibnamefont {Nussinov}}, \
  and\ \bibinfo {author} {\bibfnamefont {A.}~\bibnamefont {Seidel}},\ }\href
  {\doibase 10.1103/PhysRevLett.124.196803} {\bibfield  {journal} {\bibinfo
  {journal} {Phys. Rev. Lett.}\ }\textbf {\bibinfo {volume} {124}},\ \bibinfo
  {pages} {196803} (\bibinfo {year} {2020})}\BibitemShut {NoStop}%
\bibitem [{\citenamefont {Jain}(1989{\natexlab{a}})}]{Jain1989a}%
  \BibitemOpen
  \bibfield  {author} {\bibinfo {author} {\bibfnamefont {J.~K.}\ \bibnamefont
  {Jain}},\ }\href {\doibase 10.1103/PhysRevLett.63.199} {\bibfield  {journal}
  {\bibinfo  {journal} {Phys. Rev. Lett.}\ }\textbf {\bibinfo {volume} {63}},\
  \bibinfo {pages} {199} (\bibinfo {year} {1989}{\natexlab{a}})}\BibitemShut
  {NoStop}%
\bibitem [{\citenamefont {Jain}(1989{\natexlab{b}})}]{Jain1989}%
  \BibitemOpen
  \bibfield  {author} {\bibinfo {author} {\bibfnamefont {J.~K.}\ \bibnamefont
  {Jain}},\ }\href {\doibase 10.1103/PhysRevB.40.8079} {\bibfield  {journal}
  {\bibinfo  {journal} {Phys. Rev. B}\ }\textbf {\bibinfo {volume} {40}},\
  \bibinfo {pages} {8079} (\bibinfo {year} {1989}{\natexlab{b}})}\BibitemShut
  {NoStop}%
\bibitem [{\citenamefont {Jain}\ \emph {et~al.}(1990)\citenamefont {Jain},
  \citenamefont {Kivelson},\ and\ \citenamefont {Trivedi}}]{Jain1990}%
  \BibitemOpen
  \bibfield  {author} {\bibinfo {author} {\bibfnamefont {J.~K.}\ \bibnamefont
  {Jain}}, \bibinfo {author} {\bibfnamefont {S.~A.}\ \bibnamefont {Kivelson}},
  \ and\ \bibinfo {author} {\bibfnamefont {N.}~\bibnamefont {Trivedi}},\ }\href
  {\doibase 10.1103/PhysRevLett.64.1297} {\bibfield  {journal} {\bibinfo
  {journal} {Phys. Rev. Lett.}\ }\textbf {\bibinfo {volume} {64}},\ \bibinfo
  {pages} {1297} (\bibinfo {year} {1990})}\BibitemShut {NoStop}%
\bibitem [{\citenamefont {Jain}(1990)}]{Jain1990a}%
  \BibitemOpen
  \bibfield  {author} {\bibinfo {author} {\bibfnamefont {J.~K.}\ \bibnamefont
  {Jain}},\ }\href {\doibase 10.1103/PhysRevB.41.7653} {\bibfield  {journal}
  {\bibinfo  {journal} {Phys. Rev. B}\ }\textbf {\bibinfo {volume} {41}},\
  \bibinfo {pages} {7653} (\bibinfo {year} {1990})}\BibitemShut {NoStop}%
\bibitem [{\citenamefont {Jain}(2007)}]{jainbook}%
  \BibitemOpen
  \bibfield  {author} {\bibinfo {author} {\bibfnamefont {J.~K.}\ \bibnamefont
  {Jain}},\ }\href {\doibase 10.1017/CBO9780511607561} {\emph {\bibinfo {title}
  {{Composite Fermions}}}}\ (\bibinfo  {publisher} {Cambridge University
  Press},\ \bibinfo {year} {2007})\BibitemShut {NoStop}%
\bibitem [{\citenamefont {Jain}(2015)}]{compositeReview}%
  \BibitemOpen
  \bibfield  {author} {\bibinfo {author} {\bibfnamefont {J.~K.}\ \bibnamefont
  {Jain}},\ }\href {\doibase 10.1146/annurev-conmatphys-031214-014606}
  {\bibfield  {journal} {\bibinfo  {journal} {Annual Review of Condensed Matter
  Physics}\ }\textbf {\bibinfo {volume} {6}},\ \bibinfo {pages} {39} (\bibinfo
  {year} {2015})}\BibitemShut {NoStop}%
\bibitem [{\citenamefont {Wen}(1992)}]{Wen1992}%
  \BibitemOpen
  \bibfield  {author} {\bibinfo {author} {\bibfnamefont {X.-G.}\ \bibnamefont
  {Wen}},\ }\href {\doibase 10.1142/S0217979292000840} {\bibfield  {journal}
  {\bibinfo  {journal} {International Journal of Modern Physics B}\ }\textbf
  {\bibinfo {volume} {06}},\ \bibinfo {pages} {1711} (\bibinfo {year}
  {1992})}\BibitemShut {NoStop}%
\bibitem [{\citenamefont {Bandyopadhyay}\ \emph {et~al.}(2018)\citenamefont
  {Bandyopadhyay}, \citenamefont {Chen}, \citenamefont {Ahari}, \citenamefont
  {Ortiz}, \citenamefont {Nussinov},\ and\ \citenamefont
  {Seidel}}]{Bandyopadhyay2018}%
  \BibitemOpen
  \bibfield  {author} {\bibinfo {author} {\bibfnamefont {S.}~\bibnamefont
  {Bandyopadhyay}}, \bibinfo {author} {\bibfnamefont {L.}~\bibnamefont {Chen}},
  \bibinfo {author} {\bibfnamefont {M.~T.}\ \bibnamefont {Ahari}}, \bibinfo
  {author} {\bibfnamefont {G.}~\bibnamefont {Ortiz}}, \bibinfo {author}
  {\bibfnamefont {Z.}~\bibnamefont {Nussinov}}, \ and\ \bibinfo {author}
  {\bibfnamefont {A.}~\bibnamefont {Seidel}},\ }\href {\doibase
  10.1103/PhysRevB.98.161118} {\bibfield  {journal} {\bibinfo  {journal} {Phys.
  Rev. B}\ }\textbf {\bibinfo {volume} {98}},\ \bibinfo {pages} {161118}
  (\bibinfo {year} {2018})}\BibitemShut {NoStop}%
\bibitem [{\citenamefont {Cruise}\ and\ \citenamefont
  {Seidel}(2023)}]{Cruise2023}%
  \BibitemOpen
  \bibfield  {author} {\bibinfo {author} {\bibfnamefont {J.~R.}\ \bibnamefont
  {Cruise}}\ and\ \bibinfo {author} {\bibfnamefont {A.}~\bibnamefont
  {Seidel}},\ }\href {\doibase 10.3390/sym15020303} {\bibfield  {journal}
  {\bibinfo  {journal} {Symmetry}\ }\textbf {\bibinfo {volume} {15}} (\bibinfo
  {year} {2023}),\ 10.3390/sym15020303}\BibitemShut {NoStop}%
\bibitem [{\citenamefont {Ahari}\ \emph {et~al.}(2023)\citenamefont {Ahari},
  \citenamefont {Bandyopadhyay}, \citenamefont {Nussinov}, \citenamefont
  {Seidel},\ and\ \citenamefont {Ortiz}}]{Ahari2022}%
  \BibitemOpen
  \bibfield  {author} {\bibinfo {author} {\bibfnamefont {M.~T.}\ \bibnamefont
  {Ahari}}, \bibinfo {author} {\bibfnamefont {S.}~\bibnamefont
  {Bandyopadhyay}}, \bibinfo {author} {\bibfnamefont {Z.}~\bibnamefont
  {Nussinov}}, \bibinfo {author} {\bibfnamefont {A.}~\bibnamefont {Seidel}}, \
  and\ \bibinfo {author} {\bibfnamefont {G.}~\bibnamefont {Ortiz}},\
  }\href@noop {} {\  (\bibinfo {year} {2023})},\ \Eprint
  {http://arxiv.org/abs/2204.09684} {arXiv:2204.09684 [cond-mat.str-el]}
  \BibitemShut {NoStop}%
\bibitem [{\citenamefont {Kitaev}\ and\ \citenamefont
  {Preskill}(2006)}]{Kitaev2006}%
  \BibitemOpen
  \bibfield  {author} {\bibinfo {author} {\bibfnamefont {A.}~\bibnamefont
  {Kitaev}}\ and\ \bibinfo {author} {\bibfnamefont {J.}~\bibnamefont
  {Preskill}},\ }\href {\doibase 10.1103/PhysRevLett.96.110404} {\bibfield
  {journal} {\bibinfo  {journal} {Phys. Rev. Lett.}\ }\textbf {\bibinfo
  {volume} {96}},\ \bibinfo {pages} {2} (\bibinfo {year} {2006})}\BibitemShut
  {NoStop}%
\bibitem [{\citenamefont {Nayak}\ \emph {et~al.}(2008)\citenamefont {Nayak},
  \citenamefont {Simon}, \citenamefont {Stern}, \citenamefont {Freedman},\ and\
  \citenamefont {{Das Sarma}}}]{Nayak2008}%
  \BibitemOpen
  \bibfield  {author} {\bibinfo {author} {\bibfnamefont {C.}~\bibnamefont
  {Nayak}}, \bibinfo {author} {\bibfnamefont {S.~H.}\ \bibnamefont {Simon}},
  \bibinfo {author} {\bibfnamefont {A.}~\bibnamefont {Stern}}, \bibinfo
  {author} {\bibfnamefont {M.}~\bibnamefont {Freedman}}, \ and\ \bibinfo
  {author} {\bibfnamefont {S.}~\bibnamefont {{Das Sarma}}},\ }\href {\doibase
  10.1103/RevModPhys.80.1083} {\bibfield  {journal} {\bibinfo  {journal} {Rev.
  Mod. Phys.}\ }\textbf {\bibinfo {volume} {80}},\ \bibinfo {pages} {1083}
  (\bibinfo {year} {2008})}\BibitemShut {NoStop}%
\bibitem [{\citenamefont {Read}(2009{\natexlab{a}})}]{Read2009a}%
  \BibitemOpen
  \bibfield  {author} {\bibinfo {author} {\bibfnamefont {N.}~\bibnamefont
  {Read}},\ }\href {\doibase 10.1103/PhysRevB.79.245304} {\bibfield  {journal}
  {\bibinfo  {journal} {Phys. Rev. B}\ }\textbf {\bibinfo {volume} {79}},\
  \bibinfo {pages} {245304} (\bibinfo {year} {2009}{\natexlab{a}})}\BibitemShut
  {NoStop}%
\bibitem [{\citenamefont {Arovas}\ \emph {et~al.}(1984)\citenamefont {Arovas},
  \citenamefont {Schrieffer},\ and\ \citenamefont {Wilczek}}]{Arovas1984}%
  \BibitemOpen
  \bibfield  {author} {\bibinfo {author} {\bibfnamefont {D.}~\bibnamefont
  {Arovas}}, \bibinfo {author} {\bibfnamefont {J.~R.}\ \bibnamefont
  {Schrieffer}}, \ and\ \bibinfo {author} {\bibfnamefont {F.}~\bibnamefont
  {Wilczek}},\ }\href {\doibase 10.1103/PhysRevLett.53.722} {\bibfield
  {journal} {\bibinfo  {journal} {Phys. Rev. Lett.}\ }\textbf {\bibinfo
  {volume} {53}},\ \bibinfo {pages} {722} (\bibinfo {year} {1984})}\BibitemShut
  {NoStop}%
\bibitem [{\citenamefont {Read}(2009{\natexlab{b}})}]{Read2009}%
  \BibitemOpen
  \bibfield  {author} {\bibinfo {author} {\bibfnamefont {N.}~\bibnamefont
  {Read}},\ }\href {\doibase 10.1103/PhysRevB.79.045308} {\bibfield  {journal}
  {\bibinfo  {journal} {Phys. Rev. B}\ }\textbf {\bibinfo {volume} {79}},\
  \bibinfo {pages} {045308} (\bibinfo {year} {2009}{\natexlab{b}})}\BibitemShut
  {NoStop}%
\bibitem [{\citenamefont {Bonderson}\ \emph {et~al.}(2011)\citenamefont
  {Bonderson}, \citenamefont {Gurarie},\ and\ \citenamefont
  {Nayak}}]{Bonderson2011}%
  \BibitemOpen
  \bibfield  {author} {\bibinfo {author} {\bibfnamefont {P.}~\bibnamefont
  {Bonderson}}, \bibinfo {author} {\bibfnamefont {V.}~\bibnamefont {Gurarie}},
  \ and\ \bibinfo {author} {\bibfnamefont {C.}~\bibnamefont {Nayak}},\ }\href
  {\doibase 10.1103/PhysRevB.83.075303} {\bibfield  {journal} {\bibinfo
  {journal} {Phys. Rev. B}\ }\textbf {\bibinfo {volume} {83}},\ \bibinfo
  {pages} {075303} (\bibinfo {year} {2011})}\BibitemShut {NoStop}%
\bibitem [{\citenamefont {Flavin}\ and\ \citenamefont
  {Seidel}(2011)}]{Flavin2011}%
  \BibitemOpen
  \bibfield  {author} {\bibinfo {author} {\bibfnamefont {J.}~\bibnamefont
  {Flavin}}\ and\ \bibinfo {author} {\bibfnamefont {A.}~\bibnamefont
  {Seidel}},\ }\href {\doibase 10.1103/PhysRevX.1.021015} {\bibfield  {journal}
  {\bibinfo  {journal} {Phys. Rev. X}\ }\textbf {\bibinfo {volume} {1}},\
  \bibinfo {pages} {021015} (\bibinfo {year} {2011})}\BibitemShut {NoStop}%
\bibitem [{\citenamefont {Lee}\ and\ \citenamefont {Leinaas}(2004)}]{Lee2004}%
  \BibitemOpen
  \bibfield  {author} {\bibinfo {author} {\bibfnamefont {D.~H.}\ \bibnamefont
  {Lee}}\ and\ \bibinfo {author} {\bibfnamefont {J.~M.}\ \bibnamefont
  {Leinaas}},\ }\href {\doibase 10.1103/PhysRevLett.92.096401} {\bibfield
  {journal} {\bibinfo  {journal} {Phys. Rev. Lett.}\ }\textbf {\bibinfo
  {volume} {92}},\ \bibinfo {pages} {1} (\bibinfo {year} {2004})}\BibitemShut
  {NoStop}%
\bibitem [{\citenamefont {Seidel}\ \emph {et~al.}(2005)\citenamefont {Seidel},
  \citenamefont {Fu}, \citenamefont {Lee}, \citenamefont {Leinaas},\ and\
  \citenamefont {Moore}}]{Seidel2005}%
  \BibitemOpen
  \bibfield  {author} {\bibinfo {author} {\bibfnamefont {A.}~\bibnamefont
  {Seidel}}, \bibinfo {author} {\bibfnamefont {H.}~\bibnamefont {Fu}}, \bibinfo
  {author} {\bibfnamefont {D.-H.}\ \bibnamefont {Lee}}, \bibinfo {author}
  {\bibfnamefont {J.~M.}\ \bibnamefont {Leinaas}}, \ and\ \bibinfo {author}
  {\bibfnamefont {J.}~\bibnamefont {Moore}},\ }\href {\doibase
  10.1103/PhysRevLett.95.266405} {\bibfield  {journal} {\bibinfo  {journal}
  {Phys. Rev. Lett.}\ }\textbf {\bibinfo {volume} {95}},\ \bibinfo {pages}
  {266405} (\bibinfo {year} {2005})}\BibitemShut {NoStop}%
\bibitem [{\citenamefont {Haldane}(2011)}]{Haldane2011}%
  \BibitemOpen
  \bibfield  {author} {\bibinfo {author} {\bibfnamefont {F.~D.~M.}\
  \bibnamefont {Haldane}},\ }\href {\doibase 10.1103/PhysRevLett.107.116801}
  {\bibfield  {journal} {\bibinfo  {journal} {Phys. Rev. Lett.}\ }\textbf
  {\bibinfo {volume} {107}},\ \bibinfo {pages} {116801} (\bibinfo {year}
  {2011})}\BibitemShut {NoStop}%
\bibitem [{\citenamefont {Prange}\ \emph {et~al.}(1990)\citenamefont {Prange},
  \citenamefont {Cage}, \citenamefont {Klitzing}, \citenamefont {Girvin},
  \citenamefont {Chang}, \citenamefont {Duncan}, \citenamefont {Haldane},
  \citenamefont {Laughlin}, \citenamefont {Pruisken},\ and\ \citenamefont
  {Thouless}}]{Prange1990}%
  \BibitemOpen
  \bibfield  {author} {\bibinfo {author} {\bibfnamefont {R.~E.}\ \bibnamefont
  {Prange}}, \bibinfo {author} {\bibfnamefont {M.~E.}\ \bibnamefont {Cage}},
  \bibinfo {author} {\bibfnamefont {K.}~\bibnamefont {Klitzing}}, \bibinfo
  {author} {\bibfnamefont {S.~M.}\ \bibnamefont {Girvin}}, \bibinfo {author}
  {\bibfnamefont {A.~M.}\ \bibnamefont {Chang}}, \bibinfo {author}
  {\bibfnamefont {F.}~\bibnamefont {Duncan}}, \bibinfo {author} {\bibfnamefont
  {M.}~\bibnamefont {Haldane}}, \bibinfo {author} {\bibfnamefont {R.~B.}\
  \bibnamefont {Laughlin}}, \bibinfo {author} {\bibfnamefont {A.~M.~M.}\
  \bibnamefont {Pruisken}}, \ and\ \bibinfo {author} {\bibfnamefont {D.~J.}\
  \bibnamefont {Thouless}},\ }\href {\doibase 10.1007/978-1-4612-3350-3} {\emph
  {\bibinfo {title} {{The Quantum Hall Effect}}}},\ edited by\ \bibinfo
  {editor} {\bibfnamefont {R.~E.}\ \bibnamefont {Prange}}\ and\ \bibinfo
  {editor} {\bibfnamefont {S.~M.}\ \bibnamefont {Girvin}},\ Graduate Texts in
  Contemporary Physics\ (\bibinfo  {publisher} {Springer New York},\ \bibinfo
  {address} {New York, NY},\ \bibinfo {year} {1990})\BibitemShut {NoStop}%
\bibitem [{\citenamefont {Yang}\ \emph {et~al.}(2017)\citenamefont {Yang},
  \citenamefont {Hu}, \citenamefont {Lee},\ and\ \citenamefont
  {Papi\ifmmode~\acute{c}\else \'{c}\fi{}}}]{YangBo2017}%
  \BibitemOpen
  \bibfield  {author} {\bibinfo {author} {\bibfnamefont {B.}~\bibnamefont
  {Yang}}, \bibinfo {author} {\bibfnamefont {Z.-X.}\ \bibnamefont {Hu}},
  \bibinfo {author} {\bibfnamefont {C.~H.}\ \bibnamefont {Lee}}, \ and\
  \bibinfo {author} {\bibfnamefont {Z.}~\bibnamefont
  {Papi\ifmmode~\acute{c}\else \'{c}\fi{}}},\ }\href {\doibase
  10.1103/PhysRevLett.118.146403} {\bibfield  {journal} {\bibinfo  {journal}
  {Phys. Rev. Lett.}\ }\textbf {\bibinfo {volume} {118}},\ \bibinfo {pages}
  {146403} (\bibinfo {year} {2017})}\BibitemShut {NoStop}%
\bibitem [{\citenamefont {Yang}\ \emph {et~al.}(2012)\citenamefont {Yang},
  \citenamefont {Papi\ifmmode~\acute{c}\else \'{c}\fi{}}, \citenamefont
  {Rezayi}, \citenamefont {Bhatt},\ and\ \citenamefont {Haldane}}]{YangBo2012}%
  \BibitemOpen
  \bibfield  {author} {\bibinfo {author} {\bibfnamefont {B.}~\bibnamefont
  {Yang}}, \bibinfo {author} {\bibfnamefont {Z.}~\bibnamefont
  {Papi\ifmmode~\acute{c}\else \'{c}\fi{}}}, \bibinfo {author} {\bibfnamefont
  {E.~H.}\ \bibnamefont {Rezayi}}, \bibinfo {author} {\bibfnamefont {R.~N.}\
  \bibnamefont {Bhatt}}, \ and\ \bibinfo {author} {\bibfnamefont {F.~D.~M.}\
  \bibnamefont {Haldane}},\ }\href {\doibase 10.1103/PhysRevB.85.165318}
  {\bibfield  {journal} {\bibinfo  {journal} {Phys. Rev. B}\ }\textbf {\bibinfo
  {volume} {85}},\ \bibinfo {pages} {165318} (\bibinfo {year}
  {2012})}\BibitemShut {NoStop}%
\bibitem [{\citenamefont {Yang}(2013)}]{YangKun2013}%
  \BibitemOpen
  \bibfield  {author} {\bibinfo {author} {\bibfnamefont {K.}~\bibnamefont
  {Yang}},\ }\href {\doibase 10.1103/PhysRevB.88.241105} {\bibfield  {journal}
  {\bibinfo  {journal} {Phys. Rev. B}\ }\textbf {\bibinfo {volume} {88}},\
  \bibinfo {pages} {241105} (\bibinfo {year} {2013})}\BibitemShut {NoStop}%
\bibitem [{\citenamefont {Zhu}\ \emph {et~al.}(2017)\citenamefont {Zhu},
  \citenamefont {Sodemann}, \citenamefont {Sheng},\ and\ \citenamefont
  {Fu}}]{ZhuZheng2017}%
  \BibitemOpen
  \bibfield  {author} {\bibinfo {author} {\bibfnamefont {Z.}~\bibnamefont
  {Zhu}}, \bibinfo {author} {\bibfnamefont {I.}~\bibnamefont {Sodemann}},
  \bibinfo {author} {\bibfnamefont {D.~N.}\ \bibnamefont {Sheng}}, \ and\
  \bibinfo {author} {\bibfnamefont {L.}~\bibnamefont {Fu}},\ }\href {\doibase
  10.1103/PhysRevB.95.201116} {\bibfield  {journal} {\bibinfo  {journal} {Phys.
  Rev. B}\ }\textbf {\bibinfo {volume} {95}},\ \bibinfo {pages} {201116}
  (\bibinfo {year} {2017})}\BibitemShut {NoStop}%
\bibitem [{\citenamefont {Haldane}(2023)}]{Haldane2023}%
  \BibitemOpen
  \bibfield  {author} {\bibinfo {author} {\bibfnamefont {F.~D.~M.}\
  \bibnamefont {Haldane}},\ }\href {\doibase 10.48550/arxiv.2302.12472} {\
  (\bibinfo {year} {2023}),\ 10.48550/arxiv.2302.12472}\BibitemShut {NoStop}%
\bibitem [{\citenamefont {Ogata}\ and\ \citenamefont
  {Shiba}(1990)}]{Ogata1990}%
  \BibitemOpen
  \bibfield  {author} {\bibinfo {author} {\bibfnamefont {M.}~\bibnamefont
  {Ogata}}\ and\ \bibinfo {author} {\bibfnamefont {H.}~\bibnamefont {Shiba}},\
  }\href {\doibase 10.1103/PhysRevB.41.2326} {\bibfield  {journal} {\bibinfo
  {journal} {Phys. Rev. B}\ }\textbf {\bibinfo {volume} {41}},\ \bibinfo
  {pages} {2326} (\bibinfo {year} {1990})}\BibitemShut {NoStop}%
\bibitem [{\citenamefont {Seidel}\ and\ \citenamefont
  {Lee}(2004)}]{Seidel2004a}%
  \BibitemOpen
  \bibfield  {author} {\bibinfo {author} {\bibfnamefont {A.}~\bibnamefont
  {Seidel}}\ and\ \bibinfo {author} {\bibfnamefont {P.~A.}\ \bibnamefont
  {Lee}},\ }\href {\doibase 10.1103/PhysRevB.69.094419} {\bibfield  {journal}
  {\bibinfo  {journal} {Phys. Rev. B}\ }\textbf {\bibinfo {volume} {69}},\
  \bibinfo {pages} {1} (\bibinfo {year} {2004})}\BibitemShut {NoStop}%
\bibitem [{\citenamefont {Ribeiro}\ \emph {et~al.}(2006)\citenamefont
  {Ribeiro}, \citenamefont {Seidel}, \citenamefont {Han},\ and\ \citenamefont
  {Lee}}]{Ribeiro2006}%
  \BibitemOpen
  \bibfield  {author} {\bibinfo {author} {\bibfnamefont {T.~C.}\ \bibnamefont
  {Ribeiro}}, \bibinfo {author} {\bibfnamefont {A.}~\bibnamefont {Seidel}},
  \bibinfo {author} {\bibfnamefont {J.~H.}\ \bibnamefont {Han}}, \ and\
  \bibinfo {author} {\bibfnamefont {D.~H.}\ \bibnamefont {Lee}},\ }\href
  {\doibase 10.1209/epl/i2006-10371-6} {\bibfield  {journal} {\bibinfo
  {journal} {Europhysics Letters}\ }\textbf {\bibinfo {volume} {76}},\ \bibinfo
  {pages} {891} (\bibinfo {year} {2006})}\BibitemShut {NoStop}%
\bibitem [{\citenamefont {Kruis}\ \emph {et~al.}(2004)\citenamefont {Kruis},
  \citenamefont {McCulloch}, \citenamefont {Nussinov},\ and\ \citenamefont
  {Zaanen}}]{Kruis2004}%
  \BibitemOpen
  \bibfield  {author} {\bibinfo {author} {\bibfnamefont {H.~V.}\ \bibnamefont
  {Kruis}}, \bibinfo {author} {\bibfnamefont {I.~P.}\ \bibnamefont
  {McCulloch}}, \bibinfo {author} {\bibfnamefont {Z.}~\bibnamefont {Nussinov}},
  \ and\ \bibinfo {author} {\bibfnamefont {J.}~\bibnamefont {Zaanen}},\ }\href
  {\doibase 10.1103/PhysRevB.70.075109} {\bibfield  {journal} {\bibinfo
  {journal} {Phys. Rev. B}\ }\textbf {\bibinfo {volume} {70}},\ \bibinfo
  {pages} {075109} (\bibinfo {year} {2004})}\BibitemShut {NoStop}%
\bibitem [{\citenamefont {Chen}\ and\ \citenamefont
  {Seidel}(2014)}]{Chen2015a}%
  \BibitemOpen
  \bibfield  {author} {\bibinfo {author} {\bibfnamefont {L.}~\bibnamefont
  {Chen}}\ and\ \bibinfo {author} {\bibfnamefont {A.}~\bibnamefont {Seidel}},\
  }\href {\doibase 10.1103/PhysRevB.91.085103} {\bibfield  {journal} {\bibinfo
  {journal} {Phys. Rev. B}\ }\textbf {\bibinfo {volume} {91}},\ \bibinfo
  {pages} {085103} (\bibinfo {year} {2014})}\BibitemShut {NoStop}%
\bibitem [{\citenamefont {Mazaheri}\ \emph {et~al.}(2015)\citenamefont
  {Mazaheri}, \citenamefont {Ortiz}, \citenamefont {Nussinov},\ and\
  \citenamefont {Seidel}}]{Mazaheri2015a}%
  \BibitemOpen
  \bibfield  {author} {\bibinfo {author} {\bibfnamefont {T.}~\bibnamefont
  {Mazaheri}}, \bibinfo {author} {\bibfnamefont {G.}~\bibnamefont {Ortiz}},
  \bibinfo {author} {\bibfnamefont {Z.}~\bibnamefont {Nussinov}}, \ and\
  \bibinfo {author} {\bibfnamefont {A.}~\bibnamefont {Seidel}},\ }\href
  {\doibase 10.1103/PhysRevB.91.085115} {\bibfield  {journal} {\bibinfo
  {journal} {Phys. Rev. B}\ }\textbf {\bibinfo {volume} {91}},\ \bibinfo
  {pages} {085115} (\bibinfo {year} {2015})}\BibitemShut {NoStop}%
\bibitem [{\citenamefont {Schossler}\ \emph {et~al.}(2022)\citenamefont
  {Schossler}, \citenamefont {Bandyopadhyay},\ and\ \citenamefont
  {Seidel}}]{Schossler2022}%
  \BibitemOpen
  \bibfield  {author} {\bibinfo {author} {\bibfnamefont {M.}~\bibnamefont
  {Schossler}}, \bibinfo {author} {\bibfnamefont {S.}~\bibnamefont
  {Bandyopadhyay}}, \ and\ \bibinfo {author} {\bibfnamefont {A.}~\bibnamefont
  {Seidel}},\ }\href {\doibase 10.1103/PhysRevB.105.155124} {\bibfield
  {journal} {\bibinfo  {journal} {Phys. Rev. B}\ }\textbf {\bibinfo {volume}
  {105}},\ \bibinfo {pages} {155124} (\bibinfo {year} {2022})}\BibitemShut
  {NoStop}%
\bibitem [{\citenamefont {Chen}\ \emph {et~al.}(2019)\citenamefont {Chen},
  \citenamefont {Bandyopadhyay}, \citenamefont {Yang},\ and\ \citenamefont
  {Seidel}}]{Chen2019a}%
  \BibitemOpen
  \bibfield  {author} {\bibinfo {author} {\bibfnamefont {L.}~\bibnamefont
  {Chen}}, \bibinfo {author} {\bibfnamefont {S.}~\bibnamefont {Bandyopadhyay}},
  \bibinfo {author} {\bibfnamefont {K.}~\bibnamefont {Yang}}, \ and\ \bibinfo
  {author} {\bibfnamefont {A.}~\bibnamefont {Seidel}},\ }\href {\doibase
  10.1103/PhysRevB.100.045136} {\bibfield  {journal} {\bibinfo  {journal}
  {Phys. Rev. B}\ }\textbf {\bibinfo {volume} {100}},\ \bibinfo {pages}
  {045136} (\bibinfo {year} {2019})}\BibitemShut {NoStop}%
\bibitem [{\citenamefont {Bernevig}\ and\ \citenamefont
  {Haldane}(2007)}]{Bernevig2008}%
  \BibitemOpen
  \bibfield  {author} {\bibinfo {author} {\bibfnamefont {B.~A.}\ \bibnamefont
  {Bernevig}}\ and\ \bibinfo {author} {\bibfnamefont {F.~D.~M.}\ \bibnamefont
  {Haldane}},\ }\href {\doibase 10.1103/PhysRevLett.100.246802} {\bibfield
  {journal} {\bibinfo  {journal} {Phys. Rev. Lett.}\ }\textbf {\bibinfo
  {volume} {100}},\ \bibinfo {pages} {246802} (\bibinfo {year}
  {2007})}\BibitemShut {NoStop}%
\bibitem [{\citenamefont {Bernevig}\ and\ \citenamefont
  {Haldane}(2008)}]{Bernevig2008a}%
  \BibitemOpen
  \bibfield  {author} {\bibinfo {author} {\bibfnamefont {B.~A.}\ \bibnamefont
  {Bernevig}}\ and\ \bibinfo {author} {\bibfnamefont {F.~D.}\ \bibnamefont
  {Haldane}},\ }\href {\doibase 10.1103/PhysRevB.77.184502} {\bibfield
  {journal} {\bibinfo  {journal} {Phys. Rev. B}\ }\textbf {\bibinfo {volume}
  {77}},\ \bibinfo {pages} {1} (\bibinfo {year} {2008})}\BibitemShut {NoStop}%
\bibitem [{\citenamefont {Kl{\"u}mper}\ \emph {et~al.}(1993)\citenamefont
  {Kl{\"u}mper}, \citenamefont {Schadschneider},\ and\ \citenamefont
  {Zittartz}}]{Klumper1993}%
  \BibitemOpen
  \bibfield  {author} {\bibinfo {author} {\bibfnamefont {A.}~\bibnamefont
  {Kl{\"u}mper}}, \bibinfo {author} {\bibfnamefont {A.}~\bibnamefont
  {Schadschneider}}, \ and\ \bibinfo {author} {\bibfnamefont {J.}~\bibnamefont
  {Zittartz}},\ }\href {\doibase 10.1209/0295-5075/24/4/010} {\bibfield
  {journal} {\bibinfo  {journal} {Europhysics Letters}\ }\textbf {\bibinfo
  {volume} {24}},\ \bibinfo {pages} {293} (\bibinfo {year} {1993})}\BibitemShut
  {NoStop}%
\bibitem [{\citenamefont {Perez-Garcia}\ \emph {et~al.}(2007)\citenamefont
  {Perez-Garcia}, \citenamefont {Verstraete}, \citenamefont {Wolf},\ and\
  \citenamefont {Cirac}}]{Perez-Garcia2006}%
  \BibitemOpen
  \bibfield  {author} {\bibinfo {author} {\bibfnamefont {D.}~\bibnamefont
  {Perez-Garcia}}, \bibinfo {author} {\bibfnamefont {F.}~\bibnamefont
  {Verstraete}}, \bibinfo {author} {\bibfnamefont {M.~M.}\ \bibnamefont
  {Wolf}}, \ and\ \bibinfo {author} {\bibfnamefont {J.~I.}\ \bibnamefont
  {Cirac}},\ }\href {http://arxiv.org/abs/quant-ph/0608197} {\  (\bibinfo
  {year} {2007})},\ \Eprint {http://arxiv.org/abs/quant-ph/0608197}
  {arXiv:quant-ph/0608197} \BibitemShut {NoStop}%
\bibitem [{\citenamefont {Or\'us}(2014)}]{Orus2014}%
  \BibitemOpen
  \bibfield  {author} {\bibinfo {author} {\bibfnamefont {R.}~\bibnamefont
  {Or\'us}},\ }\href {\doibase https://doi.org/10.1016/j.aop.2014.06.013}
  {\bibfield  {journal} {\bibinfo  {journal} {Annals of Physics}\ }\textbf
  {\bibinfo {volume} {349}},\ \bibinfo {pages} {117} (\bibinfo {year}
  {2014})}\BibitemShut {NoStop}%
\bibitem [{\citenamefont {Oseledets}(2009)}]{Oseledets2009}%
  \BibitemOpen
  \bibfield  {author} {\bibinfo {author} {\bibfnamefont {I.~V.}\ \bibnamefont
  {Oseledets}},\ }\href {\doibase 10.1134/S1064562409040115} {\bibfield
  {journal} {\bibinfo  {journal} {Doklady Mathematics}\ }\textbf {\bibinfo
  {volume} {80}},\ \bibinfo {pages} {495} (\bibinfo {year} {2009})}\BibitemShut
  {NoStop}%
\bibitem [{\citenamefont {Oseledets}\ and\ \citenamefont
  {Tyrtyshnikov}(2009{\natexlab{a}})}]{Oseledets2009b}%
  \BibitemOpen
  \bibfield  {author} {\bibinfo {author} {\bibfnamefont {I.~V.}\ \bibnamefont
  {Oseledets}}\ and\ \bibinfo {author} {\bibfnamefont {E.~E.}\ \bibnamefont
  {Tyrtyshnikov}},\ }\href {\doibase 10.1137/090748330} {\bibfield  {journal}
  {\bibinfo  {journal} {SIAM Journal on Scientific Computing}\ }\textbf
  {\bibinfo {volume} {31}},\ \bibinfo {pages} {3744} (\bibinfo {year}
  {2009}{\natexlab{a}})}\BibitemShut {NoStop}%
\bibitem [{\citenamefont {Oseledets}\ and\ \citenamefont
  {Tyrtyshnikov}(2009{\natexlab{b}})}]{Oseledets2009c}%
  \BibitemOpen
  \bibfield  {author} {\bibinfo {author} {\bibfnamefont {I.~V.}\ \bibnamefont
  {Oseledets}}\ and\ \bibinfo {author} {\bibfnamefont {E.~E.}\ \bibnamefont
  {Tyrtyshnikov}},\ }\href {\doibase 10.1134/S1064562409040036} {\bibfield
  {journal} {\bibinfo  {journal} {Doklady Mathematics}\ }\textbf {\bibinfo
  {volume} {80}},\ \bibinfo {pages} {460} (\bibinfo {year}
  {2009}{\natexlab{b}})}\BibitemShut {NoStop}%
\bibitem [{\citenamefont {Bachmayr}\ \emph {et~al.}(2016)\citenamefont
  {Bachmayr}, \citenamefont {Schneider},\ and\ \citenamefont
  {Uschmajew}}]{Bachmayr2016}%
  \BibitemOpen
  \bibfield  {author} {\bibinfo {author} {\bibfnamefont {M.}~\bibnamefont
  {Bachmayr}}, \bibinfo {author} {\bibfnamefont {R.}~\bibnamefont {Schneider}},
  \ and\ \bibinfo {author} {\bibfnamefont {A.}~\bibnamefont {Uschmajew}},\
  }\href {\doibase 10.1007/s10208-016-9317-9} {\bibfield  {journal} {\bibinfo
  {journal} {Foundations of Computational Mathematics}\ }\textbf {\bibinfo
  {volume} {16}},\ \bibinfo {pages} {1423} (\bibinfo {year}
  {2016})}\BibitemShut {NoStop}%
\bibitem [{\citenamefont {Hansson}\ \emph {et~al.}(2007)\citenamefont
  {Hansson}, \citenamefont {Chang}, \citenamefont {Jain},\ and\ \citenamefont
  {Viefers}}]{Hansson2007a}%
  \BibitemOpen
  \bibfield  {author} {\bibinfo {author} {\bibfnamefont {H.}~\bibnamefont
  {Hansson}}, \bibinfo {author} {\bibfnamefont {C.-C.}\ \bibnamefont {Chang}},
  \bibinfo {author} {\bibfnamefont {J.}~\bibnamefont {Jain}}, \ and\ \bibinfo
  {author} {\bibfnamefont {S.}~\bibnamefont {Viefers}},\ }\href {\doibase
  10.1103/PhysRevB.76.075347} {\bibfield  {journal} {\bibinfo  {journal} {Phys.
  Rev. B}\ }\textbf {\bibinfo {volume} {76}},\ \bibinfo {pages} {1} (\bibinfo
  {year} {2007})}\BibitemShut {NoStop}%
\bibitem [{\citenamefont {Dubail}\ \emph {et~al.}(2012)\citenamefont {Dubail},
  \citenamefont {Read},\ and\ \citenamefont {Rezayi}}]{Dubail2012a}%
  \BibitemOpen
  \bibfield  {author} {\bibinfo {author} {\bibfnamefont {J.}~\bibnamefont
  {Dubail}}, \bibinfo {author} {\bibfnamefont {N.}~\bibnamefont {Read}}, \ and\
  \bibinfo {author} {\bibfnamefont {E.~H.}\ \bibnamefont {Rezayi}},\ }\href
  {\doibase 10.1103/PhysRevB.86.245310} {\bibfield  {journal} {\bibinfo
  {journal} {Phys. Rev. B}\ }\textbf {\bibinfo {volume} {86}},\ \bibinfo
  {pages} {245310} (\bibinfo {year} {2012})}\BibitemShut {NoStop}%
\bibitem [{\citenamefont {Zaletel}\ and\ \citenamefont
  {Mong}(2012)}]{Zaletel2012b}%
  \BibitemOpen
  \bibfield  {author} {\bibinfo {author} {\bibfnamefont {M.~P.}\ \bibnamefont
  {Zaletel}}\ and\ \bibinfo {author} {\bibfnamefont {R.~S.~K.}\ \bibnamefont
  {Mong}},\ }\href {\doibase 10.1103/PhysRevB.86.245305} {\bibfield  {journal}
  {\bibinfo  {journal} {Phys. Rev. B}\ }\textbf {\bibinfo {volume} {86}},\
  \bibinfo {pages} {245305} (\bibinfo {year} {2012})}\BibitemShut {NoStop}%
\bibitem [{\citenamefont {Tong}(2016)}]{Tong2016}%
  \BibitemOpen
  \bibfield  {author} {\bibinfo {author} {\bibfnamefont {D.}~\bibnamefont
  {Tong}},\ }\href@noop {} {\  (\bibinfo {year} {2016})},\ \Eprint
  {http://arxiv.org/abs/1606.06687} {arXiv:1606.06687 [hep-th]} \BibitemShut
  {NoStop}%
\bibitem [{\citenamefont {Estienne}\ \emph {et~al.}(2013)\citenamefont
  {Estienne}, \citenamefont {Regnault},\ and\ \citenamefont
  {Bernevig}}]{Estienne2013c}%
  \BibitemOpen
  \bibfield  {author} {\bibinfo {author} {\bibfnamefont {B.}~\bibnamefont
  {Estienne}}, \bibinfo {author} {\bibfnamefont {N.}~\bibnamefont {Regnault}},
  \ and\ \bibinfo {author} {\bibfnamefont {B.~A.}\ \bibnamefont {Bernevig}},\
  }\href@noop {} {\  (\bibinfo {year} {2013})},\ \Eprint
  {http://arxiv.org/abs/1311.2936} {arXiv:1311.2936 [cond-mat.str-el]}
  \BibitemShut {NoStop}%
\bibitem [{\citenamefont {Wu}\ \emph {et~al.}(2015)\citenamefont {Wu},
  \citenamefont {Estienne}, \citenamefont {Regnault},\ and\ \citenamefont
  {Bernevig}}]{Wu2015}%
  \BibitemOpen
  \bibfield  {author} {\bibinfo {author} {\bibfnamefont {Y.-L.}\ \bibnamefont
  {Wu}}, \bibinfo {author} {\bibfnamefont {B.}~\bibnamefont {Estienne}},
  \bibinfo {author} {\bibfnamefont {N.}~\bibnamefont {Regnault}}, \ and\
  \bibinfo {author} {\bibfnamefont {B.~A.}\ \bibnamefont {Bernevig}},\ }\href
  {\doibase 10.1103/PhysRevB.92.045109} {\bibfield  {journal} {\bibinfo
  {journal} {Phys. Rev. B}\ }\textbf {\bibinfo {volume} {92}},\ \bibinfo
  {pages} {045109} (\bibinfo {year} {2015})}\BibitemShut {NoStop}%
\bibitem [{\citenamefont {Cr{\'{e}}pel}\ \emph {et~al.}(2018)\citenamefont
  {Cr{\'{e}}pel}, \citenamefont {Estienne}, \citenamefont {Bernevig},
  \citenamefont {Lecheminant},\ and\ \citenamefont {Regnault}}]{Crepel2018}%
  \BibitemOpen
  \bibfield  {author} {\bibinfo {author} {\bibfnamefont {V.}~\bibnamefont
  {Cr{\'{e}}pel}}, \bibinfo {author} {\bibfnamefont {B.}~\bibnamefont
  {Estienne}}, \bibinfo {author} {\bibfnamefont {B.~A.}\ \bibnamefont
  {Bernevig}}, \bibinfo {author} {\bibfnamefont {P.}~\bibnamefont
  {Lecheminant}}, \ and\ \bibinfo {author} {\bibfnamefont {N.}~\bibnamefont
  {Regnault}},\ }\href {\doibase 10.1103/PhysRevB.97.165136} {\bibfield
  {journal} {\bibinfo  {journal} {Phys. Rev. B}\ }\textbf {\bibinfo {volume}
  {97}},\ \bibinfo {pages} {165136} (\bibinfo {year} {2018})}\BibitemShut
  {NoStop}%
\bibitem [{\citenamefont {Kj{\"{a}}ll}\ \emph {et~al.}(2018)\citenamefont
  {Kj{\"{a}}ll}, \citenamefont {Ardonne}, \citenamefont {Dwivedi},
  \citenamefont {Hermanns},\ and\ \citenamefont {Hansson}}]{Kjall2018}%
  \BibitemOpen
  \bibfield  {author} {\bibinfo {author} {\bibfnamefont {J.}~\bibnamefont
  {Kj{\"{a}}ll}}, \bibinfo {author} {\bibfnamefont {E.}~\bibnamefont
  {Ardonne}}, \bibinfo {author} {\bibfnamefont {V.}~\bibnamefont {Dwivedi}},
  \bibinfo {author} {\bibfnamefont {M.}~\bibnamefont {Hermanns}}, \ and\
  \bibinfo {author} {\bibfnamefont {T.~H.}\ \bibnamefont {Hansson}},\ }\href
  {\doibase 10.1088/1742-5468/aab679} {\bibfield  {journal} {\bibinfo
  {journal} {Journal of Statistical Mechanics: Theory and Experiment}\ }\textbf
  {\bibinfo {volume} {2018}},\ \bibinfo {pages} {053101} (\bibinfo {year}
  {2018})}\BibitemShut {NoStop}%
\bibitem [{\citenamefont {Affleck}\ \emph {et~al.}(1987)\citenamefont
  {Affleck}, \citenamefont {Kennedy}, \citenamefont {Lieb},\ and\ \citenamefont
  {Tasaki}}]{Affleck1987}%
  \BibitemOpen
  \bibfield  {author} {\bibinfo {author} {\bibfnamefont {I.}~\bibnamefont
  {Affleck}}, \bibinfo {author} {\bibfnamefont {T.}~\bibnamefont {Kennedy}},
  \bibinfo {author} {\bibfnamefont {E.~H.}\ \bibnamefont {Lieb}}, \ and\
  \bibinfo {author} {\bibfnamefont {H.}~\bibnamefont {Tasaki}},\ }\href
  {\doibase 10.1103/PhysRevLett.59.799} {\bibfield  {journal} {\bibinfo
  {journal} {Phys. Rev. Lett.}\ }\textbf {\bibinfo {volume} {59}},\ \bibinfo
  {pages} {799} (\bibinfo {year} {1987})}\BibitemShut {NoStop}%
\bibitem [{\citenamefont {Bagarello}(2017)}]{Bagarello2017}%
  \BibitemOpen
  \bibfield  {author} {\bibinfo {author} {\bibfnamefont {F.}~\bibnamefont
  {Bagarello}},\ }\href {\doibase 10.1134/s0040577917110083} {\bibfield
  {journal} {\bibinfo  {journal} {Theoretical and Mathematical Physics}\
  }\textbf {\bibinfo {volume} {193}},\ \bibinfo {pages} {1680} (\bibinfo {year}
  {2017})}\BibitemShut {NoStop}%
\bibitem [{\citenamefont {Zhang}\ \emph {et~al.}(2023)\citenamefont {Zhang},
  \citenamefont {Schossler}, \citenamefont {Seidel},\ and\ \citenamefont
  {Chen}}]{LiAlgebraicMR}%
  \BibitemOpen
  \bibfield  {author} {\bibinfo {author} {\bibfnamefont {F.}~\bibnamefont
  {Zhang}}, \bibinfo {author} {\bibfnamefont {M.}~\bibnamefont {Schossler}},
  \bibinfo {author} {\bibfnamefont {A.}~\bibnamefont {Seidel}}, \ and\ \bibinfo
  {author} {\bibfnamefont {L.}~\bibnamefont {Chen}},\ }\href {\doibase
  10.1103/PhysRevB.108.075142} {\bibfield  {journal} {\bibinfo  {journal}
  {Phys. Rev. B}\ }\textbf {\bibinfo {volume} {108}},\ \bibinfo {pages}
  {075142} (\bibinfo {year} {2023})}\BibitemShut {NoStop}%
\bibitem [{\citenamefont {Greiter}\ \emph {et~al.}(1992)\citenamefont
  {Greiter}, \citenamefont {Wen},\ and\ \citenamefont {Wilczek}}]{Greiter1992}%
  \BibitemOpen
  \bibfield  {author} {\bibinfo {author} {\bibfnamefont {M.}~\bibnamefont
  {Greiter}}, \bibinfo {author} {\bibfnamefont {X.-G.}\ \bibnamefont {Wen}}, \
  and\ \bibinfo {author} {\bibfnamefont {F.}~\bibnamefont {Wilczek}},\ }\href
  {\doibase 10.1016/0550-3213(92)90401-V} {\bibfield  {journal} {\bibinfo
  {journal} {Nuclear Physics B}\ }\textbf {\bibinfo {volume} {374}},\ \bibinfo
  {pages} {567} (\bibinfo {year} {1992})}\BibitemShut {NoStop}%
\bibitem [{\citenamefont {Nayak}\ and\ \citenamefont
  {Wilczek}(1996)}]{Nayak1996}%
  \BibitemOpen
  \bibfield  {author} {\bibinfo {author} {\bibfnamefont {C.}~\bibnamefont
  {Nayak}}\ and\ \bibinfo {author} {\bibfnamefont {F.}~\bibnamefont
  {Wilczek}},\ }\href {\doibase 10.1016/0550-3213(96)00430-0} {\bibfield
  {journal} {\bibinfo  {journal} {Nuclear Physics B}\ }\textbf {\bibinfo
  {volume} {479}},\ \bibinfo {pages} {529} (\bibinfo {year}
  {1996})}\BibitemShut {NoStop}%
\bibitem [{\citenamefont {Hansson}\ \emph {et~al.}(2017)\citenamefont
  {Hansson}, \citenamefont {Hermanns}, \citenamefont {Simon},\ and\
  \citenamefont {Viefers}}]{Hansson2016}%
  \BibitemOpen
  \bibfield  {author} {\bibinfo {author} {\bibfnamefont {T.~H.}\ \bibnamefont
  {Hansson}}, \bibinfo {author} {\bibfnamefont {M.}~\bibnamefont {Hermanns}},
  \bibinfo {author} {\bibfnamefont {S.~H.}\ \bibnamefont {Simon}}, \ and\
  \bibinfo {author} {\bibfnamefont {S.~F.}\ \bibnamefont {Viefers}},\ }\href
  {\doibase 10.1103/RevModPhys.89.025005} {\bibfield  {journal} {\bibinfo
  {journal} {Rev. Mod. Phys.}\ }\textbf {\bibinfo {volume} {89}},\ \bibinfo
  {pages} {025005} (\bibinfo {year} {2017})}\BibitemShut {NoStop}%
\bibitem [{\citenamefont {{Di Francesco}}\ \emph {et~al.}(1997)\citenamefont
  {{Di Francesco}}, \citenamefont {Mathieu},\ and\ \citenamefont
  {S{\'{e}}n{\'{e}}chal}}]{di1996conformal}%
  \BibitemOpen
  \bibfield  {author} {\bibinfo {author} {\bibfnamefont {P.}~\bibnamefont {{Di
  Francesco}}}, \bibinfo {author} {\bibfnamefont {P.}~\bibnamefont {Mathieu}},
  \ and\ \bibinfo {author} {\bibfnamefont {D.}~\bibnamefont
  {S{\'{e}}n{\'{e}}chal}},\ }\href {\doibase 10.1007/978-1-4612-2256-9} {\emph
  {\bibinfo {title} {{Conformal Field Theory}}}},\ Graduate Texts in
  Contemporary Physics\ (\bibinfo  {publisher} {Springer New York},\ \bibinfo
  {address} {New York, NY},\ \bibinfo {year} {1997})\BibitemShut {NoStop}%
\bibitem [{\citenamefont {Mussardo}(2009)}]{mussardo2009statistical}%
  \BibitemOpen
  \bibfield  {author} {\bibinfo {author} {\bibfnamefont {G.}~\bibnamefont
  {Mussardo}},\ }\href {https://books.google.com/books?id=JnLnXmBmCzsC} {\emph
  {\bibinfo {title} {{Statistical Field Theory: An Introduction to Exactly
  Solved Models in Statistical Physics}}}},\ Oxford Graduate Texts\ (\bibinfo
  {publisher} {OUP Oxford},\ \bibinfo {year} {2009})\BibitemShut {NoStop}%
\bibitem [{\citenamefont {Fubini}\ and\ \citenamefont
  {L{\"{u}}tken}(1991)}]{FUBINI1991}%
  \BibitemOpen
  \bibfield  {author} {\bibinfo {author} {\bibfnamefont {S.}~\bibnamefont
  {Fubini}}\ and\ \bibinfo {author} {\bibfnamefont {C.}~\bibnamefont
  {L{\"{u}}tken}},\ }\href {\doibase 10.1142/S0217732391000506} {\bibfield
  {journal} {\bibinfo  {journal} {Modern Physics Letters A}\ }\textbf {\bibinfo
  {volume} {06}},\ \bibinfo {pages} {487} (\bibinfo {year} {1991})}\BibitemShut
  {NoStop}%
\bibitem [{\citenamefont {Cristofano}\ \emph {et~al.}(1991)\citenamefont
  {Cristofano}, \citenamefont {Maiella}, \citenamefont {Musto},\ and\
  \citenamefont {Nicodemi}}]{Cristofano1991}%
  \BibitemOpen
  \bibfield  {author} {\bibinfo {author} {\bibfnamefont {G.}~\bibnamefont
  {Cristofano}}, \bibinfo {author} {\bibfnamefont {G.}~\bibnamefont {Maiella}},
  \bibinfo {author} {\bibfnamefont {R.}~\bibnamefont {Musto}}, \ and\ \bibinfo
  {author} {\bibfnamefont {F.}~\bibnamefont {Nicodemi}},\ }\href {\doibase
  10.1016/0370-2693(91)90648-A} {\bibfield  {journal} {\bibinfo  {journal}
  {Physics Letters B}\ }\textbf {\bibinfo {volume} {262}},\ \bibinfo {pages}
  {88} (\bibinfo {year} {1991})}\BibitemShut {NoStop}%
\bibitem [{\citenamefont {Milovanovi{\'{c}}}\ and\ \citenamefont
  {Read}(1996)}]{Milovanovic1996}%
  \BibitemOpen
  \bibfield  {author} {\bibinfo {author} {\bibfnamefont {M.}~\bibnamefont
  {Milovanovi{\'{c}}}}\ and\ \bibinfo {author} {\bibfnamefont {N.}~\bibnamefont
  {Read}},\ }\href {\doibase 10.1103/PhysRevB.53.13559} {\bibfield  {journal}
  {\bibinfo  {journal} {Phys. Rev. B}\ }\textbf {\bibinfo {volume} {53}},\
  \bibinfo {pages} {13559} (\bibinfo {year} {1996})}\BibitemShut {NoStop}%
\bibitem [{\citenamefont {Greiter}\ \emph {et~al.}(1991)\citenamefont
  {Greiter}, \citenamefont {Wen},\ and\ \citenamefont {Wilczek}}]{Greiter1991}%
  \BibitemOpen
  \bibfield  {author} {\bibinfo {author} {\bibfnamefont {M.}~\bibnamefont
  {Greiter}}, \bibinfo {author} {\bibfnamefont {X.-G.}\ \bibnamefont {Wen}}, \
  and\ \bibinfo {author} {\bibfnamefont {F.}~\bibnamefont {Wilczek}},\ }\href
  {\doibase 10.1103/PhysRevLett.66.3205} {\bibfield  {journal} {\bibinfo
  {journal} {Phys. Rev. Lett.}\ }\textbf {\bibinfo {volume} {66}},\ \bibinfo
  {pages} {3205} (\bibinfo {year} {1991})}\BibitemShut {NoStop}%
\bibitem [{\citenamefont {Greiter}\ and\ \citenamefont
  {Wilczek}(1992)}]{Greiter1992a}%
  \BibitemOpen
  \bibfield  {author} {\bibinfo {author} {\bibfnamefont {M.}~\bibnamefont
  {Greiter}}\ and\ \bibinfo {author} {\bibfnamefont {F.}~\bibnamefont
  {Wilczek}},\ }\href {\doibase 10.1016/0550-3213(92)90424-A} {\bibfield
  {journal} {\bibinfo  {journal} {Nuclear Physics B}\ }\textbf {\bibinfo
  {volume} {370}},\ \bibinfo {pages} {577} (\bibinfo {year}
  {1992})}\BibitemShut {NoStop}%
\bibitem [{\citenamefont {Read}\ and\ \citenamefont {Rezayi}(1996)}]{Read1996}%
  \BibitemOpen
  \bibfield  {author} {\bibinfo {author} {\bibfnamefont {N.}~\bibnamefont
  {Read}}\ and\ \bibinfo {author} {\bibfnamefont {E.}~\bibnamefont {Rezayi}},\
  }\href {\doibase 10.1103/PhysRevB.54.16864} {\bibfield  {journal} {\bibinfo
  {journal} {Phys. Rev. B}\ }\textbf {\bibinfo {volume} {54}},\ \bibinfo
  {pages} {16864} (\bibinfo {year} {1996})}\BibitemShut {NoStop}%
\bibitem [{\citenamefont {Simon}\ \emph {et~al.}(2007)\citenamefont {Simon},
  \citenamefont {Rezayi},\ and\ \citenamefont {Cooper}}]{Simon2007a}%
  \BibitemOpen
  \bibfield  {author} {\bibinfo {author} {\bibfnamefont {S.~H.}\ \bibnamefont
  {Simon}}, \bibinfo {author} {\bibfnamefont {E.~H.}\ \bibnamefont {Rezayi}}, \
  and\ \bibinfo {author} {\bibfnamefont {N.~R.}\ \bibnamefont {Cooper}},\
  }\href {\doibase 10.1103/PhysRevB.75.195306} {\bibfield  {journal} {\bibinfo
  {journal} {Phys. Rev. B}\ }\textbf {\bibinfo {volume} {75}},\ \bibinfo
  {pages} {195306} (\bibinfo {year} {2007})}\BibitemShut {NoStop}%
\bibitem [{\citenamefont {Seidel}\ and\ \citenamefont
  {Lee}(2006)}]{Seidel2006}%
  \BibitemOpen
  \bibfield  {author} {\bibinfo {author} {\bibfnamefont {A.}~\bibnamefont
  {Seidel}}\ and\ \bibinfo {author} {\bibfnamefont {D.-H.}\ \bibnamefont
  {Lee}},\ }\href {\doibase 10.1103/PhysRevLett.97.056804} {\bibfield
  {journal} {\bibinfo  {journal} {Phys. Rev. Lett.}\ }\textbf {\bibinfo
  {volume} {97}},\ \bibinfo {pages} {056804} (\bibinfo {year}
  {2006})}\BibitemShut {NoStop}%
\bibitem [{\citenamefont {Bergholtz}\ \emph {et~al.}(2006)\citenamefont
  {Bergholtz}, \citenamefont {Kailasvuori}, \citenamefont {Wikberg},
  \citenamefont {Hansson},\ and\ \citenamefont {Karlhede}}]{Bergholtz2006}%
  \BibitemOpen
  \bibfield  {author} {\bibinfo {author} {\bibfnamefont {E.~J.}\ \bibnamefont
  {Bergholtz}}, \bibinfo {author} {\bibfnamefont {J.}~\bibnamefont
  {Kailasvuori}}, \bibinfo {author} {\bibfnamefont {E.}~\bibnamefont
  {Wikberg}}, \bibinfo {author} {\bibfnamefont {T.~H.}\ \bibnamefont
  {Hansson}}, \ and\ \bibinfo {author} {\bibfnamefont {A.}~\bibnamefont
  {Karlhede}},\ }\href {\doibase 10.1103/PhysRevB.74.081308} {\bibfield
  {journal} {\bibinfo  {journal} {Phys. Rev. B}\ }\textbf {\bibinfo {volume}
  {74}},\ \bibinfo {pages} {081308} (\bibinfo {year} {2006})}\BibitemShut
  {NoStop}%
\bibitem [{\citenamefont {Weerasinghe}\ and\ \citenamefont
  {Seidel}(2014)}]{Weerasinghe2014a}%
  \BibitemOpen
  \bibfield  {author} {\bibinfo {author} {\bibfnamefont {A.}~\bibnamefont
  {Weerasinghe}}\ and\ \bibinfo {author} {\bibfnamefont {A.}~\bibnamefont
  {Seidel}},\ }\href {\doibase 10.1103/PhysRevB.90.125146} {\bibfield
  {journal} {\bibinfo  {journal} {Phys. Rev. B}\ }\textbf {\bibinfo {volume}
  {90}},\ \bibinfo {pages} {125146} (\bibinfo {year} {2014})}\BibitemShut
  {NoStop}%
\bibitem [{\citenamefont {Ortiz}\ \emph {et~al.}(2013)\citenamefont {Ortiz},
  \citenamefont {Nussinov}, \citenamefont {Dukelsky},\ and\ \citenamefont
  {Seidel}}]{Ortiz2013}%
  \BibitemOpen
  \bibfield  {author} {\bibinfo {author} {\bibfnamefont {G.}~\bibnamefont
  {Ortiz}}, \bibinfo {author} {\bibfnamefont {Z.}~\bibnamefont {Nussinov}},
  \bibinfo {author} {\bibfnamefont {J.}~\bibnamefont {Dukelsky}}, \ and\
  \bibinfo {author} {\bibfnamefont {A.}~\bibnamefont {Seidel}},\ }\href
  {\doibase 10.1103/PhysRevB.88.165303} {\bibfield  {journal} {\bibinfo
  {journal} {Phys. Rev. B}\ }\textbf {\bibinfo {volume} {88}},\ \bibinfo
  {pages} {165303} (\bibinfo {year} {2013})}\BibitemShut {NoStop}%
\bibitem [{\citenamefont {Cr\'epel}\ \emph {et~al.}(2019)\citenamefont
  {Cr\'epel}, \citenamefont {Regnault},\ and\ \citenamefont
  {Estienne}}]{Crepel2019}%
  \BibitemOpen
  \bibfield  {author} {\bibinfo {author} {\bibfnamefont {V.}~\bibnamefont
  {Cr\'epel}}, \bibinfo {author} {\bibfnamefont {N.}~\bibnamefont {Regnault}},
  \ and\ \bibinfo {author} {\bibfnamefont {B.}~\bibnamefont {Estienne}},\
  }\href {\doibase 10.1103/PhysRevB.100.125128} {\bibfield  {journal} {\bibinfo
   {journal} {Phys. Rev. B}\ }\textbf {\bibinfo {volume} {100}},\ \bibinfo
  {pages} {125128} (\bibinfo {year} {2019})}\BibitemShut {NoStop}%
\bibitem [{\citenamefont {Ruiz}(1996)}]{Ruiz1996}%
  \BibitemOpen
  \bibfield  {author} {\bibinfo {author} {\bibfnamefont {S.~M.}\ \bibnamefont
  {Ruiz}},\ }\href {http://www.jstor.org/stable/3618534} {\bibfield  {journal}
  {\bibinfo  {journal} {The Mathematical Gazette}\ }\textbf {\bibinfo {volume}
  {80}},\ \bibinfo {pages} {579} (\bibinfo {year} {1996})}\BibitemShut
  {NoStop}%
\bibitem [{\citenamefont {Balram}\ and\ \citenamefont {Jain}(2016)}]{2bar11}%
  \BibitemOpen
  \bibfield  {author} {\bibinfo {author} {\bibfnamefont {A.~C.}\ \bibnamefont
  {Balram}}\ and\ \bibinfo {author} {\bibfnamefont {J.~K.}\ \bibnamefont
  {Jain}},\ }\href {\doibase 10.1103/PhysRevB.93.235152} {\bibfield  {journal}
  {\bibinfo  {journal} {Phys. Rev. B}\ }\textbf {\bibinfo {volume} {93}},\
  \bibinfo {pages} {235152} (\bibinfo {year} {2016})}\BibitemShut {NoStop}%
\bibitem [{\citenamefont {Wu}\ \emph {et~al.}(2017)\citenamefont {Wu},
  \citenamefont {Shi},\ and\ \citenamefont {Jain}}]{wu2017new}%
  \BibitemOpen
  \bibfield  {author} {\bibinfo {author} {\bibfnamefont {Y.-H.}\ \bibnamefont
  {Wu}}, \bibinfo {author} {\bibfnamefont {T.}~\bibnamefont {Shi}}, \ and\
  \bibinfo {author} {\bibfnamefont {J.~K.}\ \bibnamefont {Jain}},\ }\href
  {\doibase 10.1021/acs.nanolett.7b01080} {\bibfield  {journal} {\bibinfo
  {journal} {Nano Letters}\ }\textbf {\bibinfo {volume} {17}},\ \bibinfo
  {pages} {4643} (\bibinfo {year} {2017})}\BibitemShut {NoStop}%
\bibitem [{\citenamefont {Balram}\ \emph
  {et~al.}(2018{\natexlab{a}})\citenamefont {Balram}, \citenamefont
  {Barkeshli},\ and\ \citenamefont {Rudner}}]{Parton_antiPfaffian}%
  \BibitemOpen
  \bibfield  {author} {\bibinfo {author} {\bibfnamefont {A.~C.}\ \bibnamefont
  {Balram}}, \bibinfo {author} {\bibfnamefont {M.}~\bibnamefont {Barkeshli}}, \
  and\ \bibinfo {author} {\bibfnamefont {M.~S.}\ \bibnamefont {Rudner}},\
  }\href {\doibase 10.1103/PhysRevB.98.035127} {\bibfield  {journal} {\bibinfo
  {journal} {Phys. Rev. B}\ }\textbf {\bibinfo {volume} {98}},\ \bibinfo
  {pages} {035127} (\bibinfo {year} {2018}{\natexlab{a}})}\BibitemShut
  {NoStop}%
\bibitem [{\citenamefont {Balram}\ \emph
  {et~al.}(2018{\natexlab{b}})\citenamefont {Balram}, \citenamefont
  {Mukherjee}, \citenamefont {Park}, \citenamefont {Barkeshli}, \citenamefont
  {Rudner},\ and\ \citenamefont {Jain}}]{1bar2bar111}%
  \BibitemOpen
  \bibfield  {author} {\bibinfo {author} {\bibfnamefont {A.~C.}\ \bibnamefont
  {Balram}}, \bibinfo {author} {\bibfnamefont {S.}~\bibnamefont {Mukherjee}},
  \bibinfo {author} {\bibfnamefont {K.}~\bibnamefont {Park}}, \bibinfo {author}
  {\bibfnamefont {M.}~\bibnamefont {Barkeshli}}, \bibinfo {author}
  {\bibfnamefont {M.~S.}\ \bibnamefont {Rudner}}, \ and\ \bibinfo {author}
  {\bibfnamefont {J.~K.}\ \bibnamefont {Jain}},\ }\href {\doibase
  10.1103/PhysRevLett.121.186601} {\bibfield  {journal} {\bibinfo  {journal}
  {Phys. Rev. Lett.}\ }\textbf {\bibinfo {volume} {121}},\ \bibinfo {pages}
  {186601} (\bibinfo {year} {2018}{\natexlab{b}})}\BibitemShut {NoStop}%
\bibitem [{\citenamefont {Faugno}\ \emph {et~al.}(2019)\citenamefont {Faugno},
  \citenamefont {Balram}, \citenamefont {Barkeshli},\ and\ \citenamefont
  {Jain}}]{fwave}%
  \BibitemOpen
  \bibfield  {author} {\bibinfo {author} {\bibfnamefont {W.~N.}\ \bibnamefont
  {Faugno}}, \bibinfo {author} {\bibfnamefont {A.~C.}\ \bibnamefont {Balram}},
  \bibinfo {author} {\bibfnamefont {M.}~\bibnamefont {Barkeshli}}, \ and\
  \bibinfo {author} {\bibfnamefont {J.~K.}\ \bibnamefont {Jain}},\ }\href
  {\doibase 10.1103/PhysRevLett.123.016802} {\bibfield  {journal} {\bibinfo
  {journal} {Phys. Rev. Lett.}\ }\textbf {\bibinfo {volume} {123}},\ \bibinfo
  {pages} {016802} (\bibinfo {year} {2019})}\BibitemShut {NoStop}%
\bibitem [{\citenamefont {Faugno}\ \emph {et~al.}(2021)\citenamefont {Faugno},
  \citenamefont {Zhao}, \citenamefont {Balram}, \citenamefont {Jolicoeur},\
  and\ \citenamefont {Jain}}]{PhysRevB.103.085303}%
  \BibitemOpen
  \bibfield  {author} {\bibinfo {author} {\bibfnamefont {W.~N.}\ \bibnamefont
  {Faugno}}, \bibinfo {author} {\bibfnamefont {T.}~\bibnamefont {Zhao}},
  \bibinfo {author} {\bibfnamefont {A.~C.}\ \bibnamefont {Balram}}, \bibinfo
  {author} {\bibfnamefont {T.}~\bibnamefont {Jolicoeur}}, \ and\ \bibinfo
  {author} {\bibfnamefont {J.~K.}\ \bibnamefont {Jain}},\ }\href {\doibase
  10.1103/PhysRevB.103.085303} {\bibfield  {journal} {\bibinfo  {journal}
  {Phys. Rev. B}\ }\textbf {\bibinfo {volume} {103}},\ \bibinfo {pages}
  {085303} (\bibinfo {year} {2021})}\BibitemShut {NoStop}%
\bibitem [{\citenamefont {Balram}\ and\ \citenamefont
  {W\'ojs}(2021)}]{PhysRevResearch.3.033087}%
  \BibitemOpen
  \bibfield  {author} {\bibinfo {author} {\bibfnamefont {A.~C.}\ \bibnamefont
  {Balram}}\ and\ \bibinfo {author} {\bibfnamefont {A.}~\bibnamefont
  {W\'ojs}},\ }\href {\doibase 10.1103/PhysRevResearch.3.033087} {\bibfield
  {journal} {\bibinfo  {journal} {Phys. Rev. Research}\ }\textbf {\bibinfo
  {volume} {3}},\ \bibinfo {pages} {033087} (\bibinfo {year}
  {2021})}\BibitemShut {NoStop}%
\bibitem [{\citenamefont {Balram}(2021)}]{PhysRevB.103.155103}%
  \BibitemOpen
  \bibfield  {author} {\bibinfo {author} {\bibfnamefont {A.~C.}\ \bibnamefont
  {Balram}},\ }\href {\doibase 10.1103/PhysRevB.103.155103} {\bibfield
  {journal} {\bibinfo  {journal} {Phys. Rev. B}\ }\textbf {\bibinfo {volume}
  {103}},\ \bibinfo {pages} {155103} (\bibinfo {year} {2021})}\BibitemShut
  {NoStop}%
\bibitem [{\citenamefont {Balram}\ \emph {et~al.}(2022)\citenamefont {Balram},
  \citenamefont {Liu}, \citenamefont {Gromov},\ and\ \citenamefont
  {Papi\ifmmode~\acute{c}\else \'{c}\fi{}}}]{Balram:2021opn}%
  \BibitemOpen
  \bibfield  {author} {\bibinfo {author} {\bibfnamefont {A.~C.}\ \bibnamefont
  {Balram}}, \bibinfo {author} {\bibfnamefont {Z.}~\bibnamefont {Liu}},
  \bibinfo {author} {\bibfnamefont {A.}~\bibnamefont {Gromov}}, \ and\ \bibinfo
  {author} {\bibfnamefont {Z.}~\bibnamefont {Papi\ifmmode~\acute{c}\else
  \'{c}\fi{}}},\ }\href {\doibase 10.1103/PhysRevX.12.021008} {\bibfield
  {journal} {\bibinfo  {journal} {Phys. Rev. X}\ }\textbf {\bibinfo {volume}
  {12}},\ \bibinfo {pages} {021008} (\bibinfo {year} {2022})}\BibitemShut
  {NoStop}%
\bibitem [{\citenamefont {Dora}\ and\ \citenamefont
  {Balram}(2022)}]{PhysRevB.105.L241403}%
  \BibitemOpen
  \bibfield  {author} {\bibinfo {author} {\bibfnamefont {R.~K.}\ \bibnamefont
  {Dora}}\ and\ \bibinfo {author} {\bibfnamefont {A.~C.}\ \bibnamefont
  {Balram}},\ }\href {\doibase 10.1103/PhysRevB.105.L241403} {\bibfield
  {journal} {\bibinfo  {journal} {Phys. Rev. B}\ }\textbf {\bibinfo {volume}
  {105}},\ \bibinfo {pages} {L241403} (\bibinfo {year} {2022})}\BibitemShut
  {NoStop}%
\end{thebibliography}%

\end{document}